\documentclass[10pt]{article}
\usepackage[section]{algorithm}
\usepackage{algpseudocode}

\usepackage{amsmath,amsfonts,amssymb}
\numberwithin{equation}{subsection}

\usepackage{pgfplotstable,filecontents,booktabs}
\usepackage{chngcntr}
\counterwithin{figure}{section}
\counterwithin{table}{section}

\usepackage{tabu}
\usepackage{longtable}
\usepackage[labelformat=simple]{subcaption}

\usepackage{tabularx}
\usepackage{verbatim}

\usepackage[OE]{express}

\newcolumntype{b}{X}
\newcolumntype{s}{>{\hsize=.4\hsize}X}
\newcolumntype{a}{>{\hsize=.2\hsize}X}

\newcommand{\lbf}[1]{\ensuremath{{\boldsymbol #1}}}

\newcommand{\for}{\text{for }}

\pgfplotstableread[col sep=comma]{average_stats.csv}\avgtable

\begin{document}
\newcommand*{\vertbar}{\rule[-1ex]{0.5pt}{2.5ex}}
\newcommand*{\horzbar}{\rule[.5ex]{2.5ex}{0.5pt}}

\title{Compressed channeled spectropolarimetry}

\author{Dennis J. Lee,\authormark{1,*} Charles F. LaCasse,\authormark{1} and Julia M. Craven\authormark{1}}

\address{\authormark{1}Sandia National Laboratories, 1515 Eubank Blvd. SE, Albuquerque, NM 87123, USA}

\email{\authormark{*}dlee1@sandia.gov}

\begin{abstract}
    Channeled spectropolarimetry measures the spectrally resolved Stokes parameters.
A key aspect of this technique is to accurately reconstruct the Stokes parameters from a modulated measurement of the channeled spectropolarimeter.
    The state-of-the-art reconstruction algorithm uses the Fourier transform to extract the Stokes parameters from channels in the Fourier domain.
While this approach is straightforward, it can be sensitive to noise and channel cross-talk, and it imposes bandwidth limitations that cut off high frequency details.
    To overcome these drawbacks, we present a reconstruction method called \emph{compressed channeled spectropolarimetry}.
    In our proposed framework, reconstruction in channeled spectropolarimetry is an underdetermined problem, where we take $N$ measurements and solve for $3 N$ unknown Stokes parameters.
    We formulate an optimization problem by creating a mathematical model of the channeled spectropolarimeter with inspiration from compressed sensing.
    We show that our approach offers greater noise robustness and reconstruction accuracy compared with the Fourier transform technique in simulations and experimental measurements.
    By demonstrating more accurate reconstructions, we push performance to the native resolution of the sensor, allowing more information to be recovered from a single measurement of a channeled spectropolarimeter.
\end{abstract}


\bibliographystyle{osajnl}

\section{Introduction}

Polarimetry and spectropolarimetry are used in a variety of applications.
Polarimetry helps to distinguish man-made targets from background clutter, evaluate stress birefringence, and characterize biological tissues \cite{Kudenov11}.
Polarization has been theorized to have applications in detecting surface features, shape, shading, and roughness \cite{Tyo06}.
It may also apply to aerosol monitoring, taking advantage of polarization dependent scatter \cite{Diner07}, and has been used for fruit quality control \cite{Boyer16}.
There is interest in employing polarization in the textile industry \cite{Peng12}.
Polarimeter design is a an active area of research \cite{Alenin14, Lowenstern16, Woodard16}, and there have been recent advances for optimal linear methods and filter design for processing channeled polarimeters \cite{LaCasse15, Alenin14}.
Spectroscopy provides insights in biomedical imaging and remote sensing \cite{Kudenov12}.
Synthesizing these capabilities, spectropolarimetry has been used to study the polarimetric and ellipsometric properties of dispersive materials \cite{Aspnes88, Oka99}.

The Stokes parameters describe incoherent, partially polarized radiation \cite{Chipman95}.
Note that we will examine the linear Stokes parameters, but the analysis in this work can be extended to include $S_3$.
There are a variety of instruments that measure the Stokes parameters \cite{Tyo06}, and we will focus on two types of channeled polarimeters.
A \emph{rotating polarizer spectropolarimeter} takes sequential measurements in time, creating channels in the temporal domain, which are combined to form an estimate of the scene.
A snapshot \emph{channeled spectropolarimeter} modulates the incident Stokes parameters onto carrier frequencies, encoding the state of polarization onto the output spectrum.
Rather than taking measurements over time, it creates channels in the spectral domain, so it requires only a single measurement.

A key aspect of this technique is to accurately reconstruct the spectrally resolved Stokes parameters from a modulated measurement of the channeled spectropolarimeter.
The state-of-the-art algorithm for reconstruction uses the Fourier transform to recover the Stokes parameters by separating them into channels based on their carrier frequencies \cite{Kudenov07}.
For convenience, we refer to this algorithm as \emph{Fourier reconstruction} (FR).
While this approach is straightforward, it suffers from noise in the measurement and from channel cross-talk.
Common experimental sources of noise include environmental vibrations, thermal fluctuations, and imperfect sampling \cite{Lee14, Lee15}.
In addition, Fourier reconstruction imposes bandwidth limitations from windowing the Fourier transform in order to filter out channels, thus cutting off high frequency details \cite{LaCasse11}.

To overcome these drawbacks, we propose a reconstruction method called \emph{compressed channeled spectropolarimetry} (CCSP).
    In our proposed framework, reconstruction in channeled spectropolarimetry is an underdetermined problem, where we take $N$ measurements and solve for $3 N$ unknown Stokes parameters.
    We formulate an optimization problem by creating a mathematical model of the channeled spectropolarimeter with inspiration from compressed sensing \cite{Donoho06, LeeSPIE16, LeeArxiv16}.
    We demonstrate that our approach offers greater noise robustness and reconstruction accuracy compared with Fourier reconstruction in simulations and experimental measurements.
    It reduces the need for windowing used in Fourier reconstruction to extract channels.
    We will consider 1D signals, but our analysis can be extended to higher dimensional data such as images by vectorizing the relevant quantities or by processing the data in one dimensional scans.
    More generally, our analysis applies to all channeled polarimeters, including those that are temporally or spatially channeled, by solving for the Stokes parameters from a system of underdetermined equations.
    Our framework enables future research to reconstruct Stokes parameters with less than $N$ measurements while maintaining the same resolution, potentially allowing sensors to be smaller in size, lighter weight, and lower power.

\section{Theory}\label{SectionTheory}

In this section we will describe how to reconstruct Stokes parameters from a rotating polarizer spectropolarimeter.
Then we will introduce a model of the channeled spectropolarimeter, which requires only one measurement.
We will review the state-of-the-art algorithm for recovering Stokes parameters from the channeled spectropolarimeter based on the Fourier transform.
Finally we will present compressed channeled spectropolarimetry, a framework for reconstructing the Stokes parameters from an underdetermined system.

\subsection{Reconstruction from a rotating polarizer spectropolarimeter} \label{SectionRotatingPolarizer}
A rotating polarizer spectropolarimeter consists of a polarizer followed by a spectrometer.
This instrument takes multiple measurements by rotating the polarizer in steps, and the spectrally resolved Stokes parameters can be recovered from these measurements.
We define some notation for a mathematical description of the spectropolarimeter.
The Stokes parameters can be expressed in terms of intensities measured by linear polarizers:
    \begin{equation}
        \left[ \begin{array}{c}
            S_0 \\
            S_1 \\
            S_2 \\
            S_3 \\
        \end{array} \right]
         = \left[ \begin{array}{c}
            I_H + I_V \\
            I_H - I_V \\
             I_{45} - I_{135} \\
            I_L - I_R \\
        \end{array} \right]
    \end{equation}
    where $I_H$, $I_V$, $I_{45}$, $I_{135}$, $I_L$, and $I_R$ are the intensities observed through horizontal, vertical, $45^\circ$, $135^\circ$, left-circular, and right-circular polarizers, respectively \cite{Chipman95}.
    In this work, we will consider the linear polarization states, but our analysis can be extended to include $S_3$.

Let $\lbf s_0$, $\lbf s_1$, and $\lbf s_2$ be the spectrally resolved Stokes parameters:
    \begin{equation}
        \begin{array}{cccc}
            \lbf s_0 = \bigl[S_0(\sigma_1) & \ldots & S_0(\sigma_N)\bigr]^T
        \end{array} \in \mathbb R^{N},
        \label{s0def}
    \end{equation}
    \begin{equation}
        \begin{array}{cccc}
            \lbf s_1 = \bigl[S_1(\sigma_1) & \ldots & S_1(\sigma_N)\bigr]^T
        \end{array} \in \mathbb R^{N},
        \label{s1def}
    \end{equation}
and
    \begin{equation}
        \begin{array}{cccc}
            \lbf s_2 = \bigl[S_2(\sigma_1) & \ldots & S_2(\sigma_N)\bigr]^T
        \end{array} \in \mathbb R^{N}
        \label{s2def}
    \end{equation}
where $N$ is the number of wavenumbers, and $\sigma$ is the variable for wavenumber.
Boldface variables denote \emph{mathematical} vectors (not to be confused with Stokes vectors).
We concatenate the Stokes parameters into a matrix $\lbf S$:
    \begingroup
    \renewcommand*{\arraystretch}{1.5}
    \begin{equation}
        \lbf S = \left[ \begin{array}{ccc}
            \horzbar & \lbf s_0^T & \horzbar \\
            \horzbar & \lbf s_1^T & \horzbar \\
            \horzbar & \lbf s_2^T & \horzbar \\
        \end{array} \right] \in \mathbb R^{3 \times N}.
    \end{equation}
    \endgroup
Let $\lbf \Theta$ be a matrix with each row as the analyzer vector for a polarizer oriented at angle $\theta_i$:
    \begin{equation}
        \lbf \Theta = \left[ \begin{array}{ccc}
            1 & cos(2 \theta_1) & sin(2 \theta_1) \\
              & \vdots & \\
            1 & cos(2 \theta_A) & sin(2 \theta_A) \\
        \end{array} \right] \in \mathbb R^{A \times 3}
    \end{equation}
    where $A$ is the number of angles.
Let $\lbf Y$ be a matrix of measurements such that $\lbf Y = \lbf \Theta \lbf S$:
    \begin{equation}
        \lbf Y = \left[ \begin{array}{ccc}
            \horzbar & \lbf y_1 & \horzbar \\
              & \vdots & \\
            \horzbar & \lbf y_A & \horzbar \\
        \end{array} \right] = \lbf \Theta \lbf S \in \mathbb R^{A \times N}
    \end{equation}
    where $\lbf y_i \in \mathbb R^N$ is the measured spectrum when the linear polarizer is oriented at angle $\theta_i$.
We solve for the spectrally resolved Stokes parameters as
    \begin{equation}
        \lbf S = \lbf \Theta^{-1} \lbf Y
    \end{equation}
given $A = 3$ measurements.
If the polarizer is stepped through more than $A = 3$ angles, we can estimate the Stokes parameters with a least squares fit to account for noise:
    \begin{equation}
        \lbf S = \left(\lbf \Theta^T \lbf \Theta \right)^{-1} \lbf \Theta^T \lbf Y.
    \end{equation}

\subsection{Optical system of a channeled spectropolarimeter}
A channeled spectropolarimeter requires only a single measurement, unlike the rotating polarizer spectropolarimeter.
Our optical system consists of a quarter wave plate (QWP), a retarder (R) oriented at $45^\circ$, and a horizontally oriented polarizer (LP).
These elements are described by the Mueller matrices
\begin{equation}
\lbf{M}_{\text{QWP}}= \left[
\begin{array}{cccc}
1 & 0 & 0 & 0\\
0 & 1 & 0 & 0\\
0 & 0 & 0 & 1\\
0 & 0 & -1 & 0
\end{array}
\right],
\end{equation}
\begin{equation}
\lbf{M}_{R}(\sigma) = \left[
\begin{array}{cccc}
1 & 0 & 0 & 0\\
    0 & \text{cos}[\phi(\sigma)] & 0 & -\text{sin}[\phi(\sigma)]\\
0 & 0 & 1 & 0\\
0 & \text{sin}[\phi(\sigma)] & 0 & \text{cos}[\phi(\sigma)]
\end{array}
\right],
\end{equation}
and
\begin{equation}
    \lbf{M}_{\text{LP}} = \frac{1}{2} \left[
\begin{array}{cccc}
1 & 1 & 0 & 0\\
1 & 1 & 0 & 0\\
0 & 0 & 0 & 0\\
0 & 0 & 0 & 0
\end{array}
\right].
\end{equation}
Let $\lbf s_{\text{in}}(\sigma)$ be the input Stokes vector:
\begin{equation}
\lbf s_{\text{in}}(\sigma) = \left[
\begin{array}{cccc}
S_0(\sigma)
 & S_1(\sigma)
 & S_2(\sigma)
& S_3(\sigma)
\end{array}
\right]^T,
\end{equation}
and let $\lbf s_{\text{out}}(\sigma)$ be the output Stokes vector of the optical system:
\begin{equation}
\lbf s_{\text{out}}(\sigma) = \lbf{M}_{\text{LP}}
~\lbf{M}_{R}(\sigma)
~\lbf{M}_{\text{QWP}} ~\lbf s_{\text{in}}(\sigma).
\end{equation}
There are other possible variations on the channeled spectropolarimeter design.
For example, the quarter wave plate may be replaced with another higher order retarder to measure $S_3(\sigma)$ \cite{Oka99}.
The analysis in this work can be extended to other configurations by using the appropriate Mueller matrices to model the system.

After light passes through the optical elements, the spectrometer measures the intensity as
\begin{equation}
    \begin{split}
        y(\sigma) &= \frac{1}{2}\left\{ S_0(\sigma) + S_1(\sigma) \text{cos}[\phi(\sigma)] + S_2(\sigma) \text{sin}[\phi(\sigma)] \right\} \\
        &= \frac{1}{2} \left\{ S_0(\sigma) + S_1(\sigma) \text{cos}[2 \pi B(\sigma) \, t \sigma] + S_2(\sigma) \text{sin}[2 \pi B(\sigma) \, t \sigma] \right\} \\
        &= \frac{1}{2} \left\{ S_0(\sigma) + S_1(\sigma) \text{cos}(2 \pi f_c \sigma) + S_2(\sigma) \text{sin}(2 \pi f_c \sigma) \right\}
    \end{split}
    \label{eqn:ysigma}
\end{equation}
where $\phi(\sigma)$ is the phase of the optical system
\begin{equation}
\phi(\sigma) = 2 \pi B(\sigma) \, t \sigma
\label{eqn:phi}
\end{equation}
with $ \sigma = 1/\lambda $, $ B(\sigma) = | n_o(\sigma) -
n_e(\sigma) | $, and $t$ is the thickness of the retarder. The carrier frequency is
\begin{equation}
    f_c = B(\sigma) \, t.
    \label{fc}
\end{equation}

\subsection{Fourier reconstruction}\label{sectionFR}

The state-of-the-art algorithm for reconstruction uses the Fourier transform to recover the Stokes parameters by separating them into channels based on the carrier frequencies of the output spectrum $y(\sigma)$ \cite{Kudenov07}.
For convenience, we refer to this approach as \emph{Fourier reconstruction} (FR).
The first step is to take the inverse Fourier transform of the output spectrum to obtain an interferogram:
\begin{equation}
    \begin{split}
        \widehat{y}(d) &= \mathcal F^{-1} [y(\sigma)] \\
        &= \frac{1}{2} \widehat{S}_0(d) + \frac{1}{4} \left[\widehat{S}_1(d - Bt) + \widehat{S}_1(d + Bt)\right] + \frac{1}{4} i \left[\widehat{S}_2(d + Bt) - \widehat{S}_2(d - Bt)\right]
    \end{split}
\end{equation}
where $\widehat{y}(d)$ is the inverse Fourier transform of $y(\sigma)$, and $d$ is an optical path difference variable.
To simplify our analysis, we assume that birefringence has negligible variation with wavenumber.

The next step is to filter the interferogram to isolate the channel centered at zero,
\begin{equation}
    \begin{split}
        \widehat{C}_0(d) &= 2 \cdot H_{\text{LPF}}(d) \cdot \widehat{y}(d) \\
        &= \widehat{S}_0(d),
    \end{split}
    \label{eqn:frC0}
\end{equation}
and the channel with a peak at carrier frequency $f_c = Bt$,
\begin{equation}
    \begin{split}
        \widehat{C}_1(d) &= H_{\text{BPF}}(d) \cdot \widehat{y}(d) \\
        &= \frac{1}{4} \left[\widehat{S}_1(d + Bt) + i\widehat{S}_2(d + Bt)\right]
    \end{split}
    \label{eqn:frC1}
\end{equation}
where $H_{\text{LPF}}(d)$ is a lowpass filter and $H_{\text{BPF}}(d)$ is a bandpass filter to isolate the sideband at $d = Bt$.
Some examples of the filters $H_{\text{LPF}}(d)$ and $H_{\text{BPF}}(d)$ include common functions such as rectangular, Hamming, or Blackman windows.
The filter center of $H_{\text{BPF}}(d)$ is centered at the peak of the sideband, which is typically near or at the carrier frequency, and the filter widths of $H_{\text{LPF}}(d)$ and $H_{\text{BPF}}(d)$ are commonly chosen to be the same to maintain equal spectral resolution in both channels.

The third step is to take the Fourier transform of the channels:
\begin{equation}
    \begin{split}
        C_0(\sigma) &= \mathcal F \left[ \widehat{C}_0(d) \right] \\
        &= S_0(\sigma)
    \end{split}
    \label{eqn:c0s0}
\end{equation}
and
\begin{equation}
    \begin{split}
    C_1(\sigma) &= \mathcal F \left[ \widehat{C}_1(d) \right] \\
        &= \frac{1}{4} \left[ S_1(\sigma) e^{i2\pi \sigma B t} + i S_2(\sigma) e^{i 2\pi \sigma B t} \right].
    \end{split}
    \label{eqn:c1}
\end{equation}

We can estimate the phase of the optical system, $\phi(\sigma) = 2 \pi \sigma B t$, by taking a reference measurement.
For the reference measurement, the channeled spectropolarimeter measures a horizontal polarizer with $S_1^R(\sigma) / S_0^R(\sigma) = 1$ and $S_2^R(\sigma) = 0$.
Note that other reference samples are possible, such as a vertical polarizer.
We can write expressions for $C_0^R(\sigma)$ and $C_1^R(\sigma)$ based on the same analysis used to determine $C_0(\sigma)$ and $C_1(\sigma)$:
\begin{equation}
    \begin{split}
        C_0^R(\sigma) &= \mathcal F \left[ \widehat{C}_0^R(d) \right] \\
        &= S_0^R(\sigma)
    \end{split}
    \label{eqn:c0Rs0R}
\end{equation}
and
\begin{equation}
    C_1^R(\sigma) = \frac{1}{4} S_1^R(\sigma) e^{i 2 \pi \sigma B t}.
\end{equation}
The phase of the optical system, $\phi(\sigma) = 2 \pi \sigma B t$, can be estimated from $C_1^R(\sigma)$:
\begin{equation}
    \widehat{\phi}(\sigma) = \text{arg} \left[ C_1^R(\sigma) \right].
    \label{eqn:phihat}
\end{equation}
Further manipulation helps to isolate $S_1(\sigma)$ and $S_2(\sigma)$:
\begin{equation}
    \begin{split}
        \overline{C}_1(\sigma) &= \frac{C_1(\sigma)}{C_1^R(\sigma)} \cdot \frac{S_0^R(\sigma)}{S_0(\sigma)} \\
        &= \frac{S_1(\sigma) / S_0(\sigma)}{S_1^R(\sigma) / S_0^R(\sigma)} + i \frac{S_2(\sigma) / S_0(\sigma)}{S_1^R(\sigma) / S_0^R(\sigma)} \\
        &= \frac{S_1(\sigma)}{S_0(\sigma)} + i \frac{S_2(\sigma)}{S_0(\sigma)}
    \end{split}
\end{equation}
where $S_0(\sigma) = C_0(\sigma)$ and $S_0^R(\sigma) = C_0^R(\sigma)$ from Eqs. (\ref{eqn:c0s0}) and (\ref{eqn:c0Rs0R}).
Finally, we extract $S_1(\sigma)/S_0(\sigma)$ and $S_2(\sigma)/S_0(\sigma)$ from $\overline{C}_1(\sigma)$:
\begin{equation}
    \frac{S_1(\sigma)}{S_0(\sigma)} = \text{Re} \left[ \overline{C}_1(\sigma) \right]
    \label{eqn:s1fr}
\end{equation}
and
\begin{equation}
    \frac{S_2(\sigma)}{S_0(\sigma)} = \text{Im} \left[ \overline{C}_1(\sigma) \right].
    \label{eqn:s2fr}
\end{equation}
Note that the recovered Stokes parameters from Eqs. (\ref{eqn:c0s0}) and (\ref{eqn:s1fr})--(\ref{eqn:s2fr}) are estimates.

While this approach is straightforward, it suffers from noise in the measurement and from channel cross-talk.
It also requires the choice of a window function to extract the channels $\widehat{C}_0(d)$ and $\widehat{C}_1(d)$ as described in Eqs. (\ref{eqn:frC0})--(\ref{eqn:frC1}).
The window imposes bandwidth limitations, which cuts off high frequency details.

\subsection{Compressed channeled spectropolarimetry}\label{sectionCS}

To overcome the drawbacks of Fourier reconstruction, we propose a reconstruction method called \emph{compressed channeled spectropolarimetry} (CCSP).
    In our proposed framework, reconstruction in channeled spectropolarimetry is an underdetermined problem, where we take $N$ measurements and solve for $3 N$ unknown Stokes parameters.
    In this section we formulate an optimization problem by creating a mathematical model of the channeled spectropolarimeter with inspiration from compressed sensing.

Let $\lbf s$ be a Stokes vector:
    \begin{align}
        \lbf s &= \begin{bmatrix}
              \lbf s_{0} \\
               \lbf s_{1} \\
               \lbf s_{2}
        \end{bmatrix} \in \mathbb R^{3 N}.
        \label{StokesVector}
    \end{align}
where the Stokes parameters $\lbf s_{0}$, $\lbf s_{1}$, and $\lbf s_{2}$ are defined in Eqs. (\ref{s0def})--(\ref{s2def}).
The boldface notation denotes mathematical vectors, where $\lbf s_i \in \mathbb R^N$ for $i = 0, 1, 2$.
Let us define two diagonal matrices:
    \begin{equation}
        \lbf M_\text{cos} = \text{diag}[\text{cos}(\phi_1), \ldots, \text{cos}(\phi_N)] \in \mathbb R^{N \times N}
    \end{equation}
and
    \begin{equation}
        \lbf M_\text{sin} = \text{diag}[\text{sin}(\phi_1), \ldots, \text{sin}(\phi_N)] \in \mathbb R^{N \times N}.
    \end{equation}
    The phase vector
\begin{equation}
\lbf \phi = \left[
\begin{array}{ccc}
\phi_1
 & \dots
& \phi_N
\end{array}
\right]^T
\label{eqn:PhaseVector}
\end{equation}
       can be estimated from a reference measurement of a horizontal polarizer.
       Equation (\ref{eqn:phihat}) describes how to estimate the phase of the optical system, $\widehat{\phi}(\sigma)$.
       Let us define a model matrix $\lbf M_{\text{model}}$ using the matrices above:
    \begin{equation}
        \lbf M_{\text{model}} = \frac{1}{2} \left[ \begin{array}{c|c|c}
              \lbf I & \lbf M_\text{cos} & \lbf M_\text{sin}
        \end{array} \right] \in \mathbb R^{N \times 3 N}
        \label{Mcsp}
    \end{equation}
where $ \lbf I \in \mathbb R^{N \times N} $ is the identity matrix.
The output of a channeled spectropolarimeter is a spectrum that encodes the state of polarization.
Using the notation above, we express the output spectrum as
    \begin{equation}
        \lbf y_\text{model} = \lbf M_\text{model} \lbf s,
    \end{equation}
with $i$th entry
    \begin{equation}
        y_{\text{model, i}} = \frac{1}{2} \left[S_0(\sigma_i) + S_1(\sigma_i) \text{cos}(\phi_i) + S_2(\sigma_i) \text{sin}(\phi_i) \right].
        \label{eqn:ymodeli}
    \end{equation}

    We will represent the Stokes vector $\lbf s$ in terms of coefficients from discrete cosine transform (DCT) bases and Legendre polynomials.
    The DCT coefficients help to capture sinusoidal variations.
    However, the DCT does not compactly represent low order polynomials.
    The Legendre polynomials are an orthogonal basis that help to model signals such as linear, quadratric, and cubic polynomials.
Let $\lbf p_n$ be the $n$th polynomial basis vector:
    \begin{equation}
        \begin{array}{cccc}
            \lbf p_n = \bigl[P_n(x_1) & \ldots & P_n(x_N)\bigr]^T
        \end{array} \in \mathbb R^{N}
    \end{equation}
where $x_1, \ldots, x_N$ uniformly sample the interval $[-1, 1]$, and the Legendre polynomial $P_n(x)$ is
    \begin{equation}
        P_n(x) = 2^n \sum_{k = 0}^n x^k \binom{n}{k} \binom{\frac{n + k - 1}{2}}{n}.
    \end{equation}
Let $\lbf M_{\text{dct}}$ be a DCT matrix with $(m, n)$th entry
    \begin{equation}
        \lbf M_{\text{dct}}(m, n) = \begin{cases}
            \sqrt{\frac{2}{N}} \text{cos} \left( \frac{\pi}{2 N} (2 n - 1) (m - 1) \right), & \for m = 2, \ldots, N \\
            \sqrt{\frac{1}{N}}, & \for m = 1
        \end{cases}, ~\lbf M_{\text{dct}} \in \mathbb R^{N \times N}.
    \end{equation}
We combine the polynomial and DCT bases in a $N \times (N + L)$  support matrix $\lbf M_{\text{support}}^N$:
    \begin{equation}
        \lbf M_{\text{support}}^N = \left[ \begin{array}{ccc|c}
            \vertbar &  & \vertbar & \\
            \lbf p_1 & \ldots & \lbf p_L & \lbf M_{\text{dct}} \\
            \vertbar &  & \vertbar & \\
        \end{array} \right] \in \mathbb R^{N \times (N + L)}
    \end{equation}
where $L$ is the number of Legendre polynomials.

The Stokes parameters can be recovered from their basis coefficients by
    \begin{equation}
            \lbf s_i = \lbf M_{\text{support}}^N ~\widehat{\lbf s}_i,
            \label{eqn:RecoverStokesi}
    \end{equation}
where $ \widehat{\lbf s}_i \in \mathbb R^{N + L}$ are the basis coefficients for Stokes parameters $ \lbf s_i $, $i = 0, 1, 2$ from Eqs. (\ref{s0def})--(\ref{s2def}).

The basis coefficients represent both the DCT and Legendre polynomials, and we can label the basis associated with each coefficient as
    \begingroup
    \renewcommand*{\arraystretch}{1.5}
    \begin{align}
        \widehat{\lbf s}_i &= \begin{bmatrix}
            \widehat{\lbf s}_i^{\text{ poly}} \\
               \widehat{\lbf s}_i^{\text{ DCT}}
        \end{bmatrix}
    \end{align}
    \endgroup
for $i = 0, 1, 2$. The basis coefficients that represent Legendre polynomials and the DCT are $\widehat{\lbf s}_i^{\text{ poly}} \in \mathbb R^{L}$ and $\widehat{\lbf s}_i^{\text{ DCT}} \in \mathbb R^{N}$, respectively.

Let $\widehat{\lbf s}$ be a concatenation of basis coefficients $\widehat{\lbf s_i}$:
    \begin{align}
        \widehat{\lbf s} &= \begin{bmatrix}
               \widehat{\lbf s}_0 \\
               \widehat{\lbf s}_1 \\
               \widehat{\lbf s}_2
        \end{bmatrix} \quad \in \mathbb R^{3 (N + L)}.
    \end{align}
The Stokes vector $\lbf s$ can be recovered from its basis coefficents $\widehat{\lbf s}$ by
    \begin{equation}
            \lbf s = \lbf M_{\text{support}} ~\widehat{\lbf s},
            \label{eqn:RecoverStokes}
    \end{equation}
where the support matrix $\lbf M_{\text{support}}$ is built from $\lbf M_{\text{support}}^N$:
    \begingroup
    \renewcommand*{\arraystretch}{1.5}
    \begin{equation}
        \lbf M_{\text{support}} = \left[ \begin{array}{c|c|c}
            \lbf M_{\text{support}}^N & \lbf{0} & \lbf{0} \\ \hline
            \lbf{0} & \lbf M_{\text{support}}^N & \lbf{0} \\ \hline
            \lbf{0} & \lbf{0} & \lbf M_{\text{support}}^N
        \end{array} \right] \in \mathbb R^{3 N \times 3 (N + L)}.
        \label{Mbasis}
    \end{equation}
    \endgroup
Here $\lbf 0 \in \mathbb R^{N \times (N + L)}$ is a matrix of zeroes.
We wish to represent the output of the channeled spectropolarimeter in terms of the basis coefficients using a matrix $\lbf A$:
    \begin{equation}
        \lbf y_{\text{model}} = \lbf A \widehat{\lbf s}
    \end{equation}
where $\lbf A$ is the reconstruction matrix,
    \begin{equation}
        \lbf A = \lbf M_{\text{model}} \cdot \lbf M_{\text{support}} \quad \in \mathbb R^{N \times 3(N + L)}.
    \end{equation}
Our goal is to solve an optimization problem for the basis coefficients $\widehat{\lbf s}$:
    \begin{equation}
        \begin{aligned}
            & \underset{\widehat{\lbf s}}{\text{minimize}}
            & & || \lbf A \widehat{\lbf s} - \lbf y ||^2_2 + \beta \left(|| \widehat{\lbf s}_0 ||_1 + || \widehat{\lbf s}_1 ||_1 + || \widehat{\lbf s}_2 ||_1\right)\\
            & \text{subject to}
            & & \widehat{\lbf s}_i^{\text{ DCT}}(f) = 0, \quad f \geq \tau, \quad i = 0, 1, 2
        \end{aligned}
        \label{MinimizationProblem}
    \end{equation}
where $\lbf y \in \mathbb R^N$ is the measured spectrum, and $\tau$ is a frequency threshold to help suppress oscillatory artifacts.
The likelihood term $|| \lbf A \widehat{\lbf s} - \lbf y ||^2_2$ minimizes the error with measured data.
The regularizer term $\beta \left(|| \widehat{\lbf s}_0 ||_1 + || \widehat{\lbf s}_1 ||_1 + || \widehat{\lbf s}_2 ||_1\right)$ contains $L_1$ norms which promote sparsity in the basis coefficients $\widehat{\lbf s}_0$, $\widehat{\lbf s}_1$, $\widehat{\lbf s}_2$.
When the regularizer weight $\beta$ is increased, the solution favors more sparse solutions.
As the signal-to-noise ratio (SNR) decreases, increasing $\beta$ helps to improve robustness to noise, as we will discuss in Section \ref{SectionSNR}.

The constraint $\widehat{\lbf s}_i^{\text{ DCT}}(f) = 0$ for $f \geq \tau$ and $i = 0, 1, 2$ sets high frequency DCT coefficients to zero according to a threshold $\tau$.
This constraint provides the option to set the threshold $\tau$ to suppress oscillatory artifacts.
For example, it may be known that the Stokes parameters contain negligible frequency content above the carrier frequency.
In this case, a user could choose $\tau$ such that it is close to the carrier frequency.
Other types of \emph{a priori} knowledge could be implemented similarly.
In comparison, Fourier reconstruction imposes windowing functions that cut off channel frequencies at \emph{half} of the carrier frequency to maintain equal spectral resolution in both channels.
This guideline for setting $\tau$ doubles the spectral content of the Stokes parameters compared to Fourier reconstruction.
Alternatively, the threshold $\tau$ can be set to a high value, or the constraint can be removed.
If the reconstruction exhibits high frequency oscillations, $\tau$ can be decreased and tuned to remove the oscillations.

We describe the processing steps of the algorithm as follows.
Problem (\ref{MinimizationProblem}) requires a few inputs. One input is the spectrometer measurement $\lbf y$.
Other inputs are the estimated phase of the optical system, $\widehat{\phi}(\sigma)$, described in Eq. (\ref{eqn:phihat}), and the matrices $\lbf M_{\text{model}}$, $\lbf M_{\text{support}}$, and $\lbf A$.
The final inputs are the reconstruction parameters $\beta$ and $\tau$.
To solve Problem (\ref{MinimizationProblem}), we note that it is a second order cone program \cite{Boyd04}, and interior point methods are widely used to solve these convex optimization problems \cite{Boyd04, Wright97}.
We use an interior point solver for second order cone programs called the embedded conic solver \cite{Domahidi13}.
The output of the optimization is the solution $\widehat{\lbf s}$,
and the Stokes parameters $\lbf s$ can be recovered from their coefficients $\widehat{\lbf s}$ via Eq. (\ref{eqn:RecoverStokes}).

    Our framework enables future research to reconstruct Stokes parameters with less than $N$ measurements while maintaining the same resolution, potentially allowing sensors to be smaller in size, lighter weight, and lower power.
For example, given $M < N$ measurements, an interesting question is how well we can reconstruct $3 N$ Stokes parameters.
Alternatively, can we increase the resolution by reconstructing more than $3 N$ parameters given $N$ measurements?
By formulating reconstruction as solving an underdetermined system, we open the avenue for research into these questions and push performance to the native resolution of the sensor by recovering more information from a single measurement of a channeled spectropolarimeter.

\section{Simulation}\label{SectionSimulation}

In this section we will simulate measurements from a channeled spectropolarimeter.
First we will reconstruct Stokes parameters under varying noise, and then we will vary the frequency of the Stokes parameters.
We will present metrics for measuring how well the reconstructions fit ground truth and compare algorithm performance.

\subsection{Test cases with varying noise}\label{SectionSNR}

Our goal in this section is to reconstruct Stokes parameters with varying noise.
We simulate a reference measurement by setting $S_0^R(\sigma) = S_1^R(\sigma) = 1$ and $S_2^R(\sigma) = 0$.
As described in Eq. (\ref{eqn:phihat}), we produce an estimated phase of the optical sytem, $\widehat{\phi}(\sigma)$, from the reference measurement.
For the sample output, we will set $S_0(\sigma)$ to be a cosine with frequency $f_\text{S0}$:
\begin{equation}
    S_0(\sigma) = a_1 \text{cos} (2 \pi f_\text{S0} \sigma) + b
    \label{eqn:s0snr}
\end{equation}
with
   \begin{equation}
    S_1(\sigma) = a_2 S_0(\sigma)
    \label{eqn:s1snr}
\end{equation}
and
\begin{equation}
    S_2(\sigma) = 0,
    \label{eqn:s2snr}
\end{equation}
 where $a_1$ and $a_2$ are attenuation factors and $b$ is an offset.
  In our simulations, $a_1 = 0.8$, $a_2 = 0.5$, and $b = 2.5$, and we set the ratio $f_{S0}/f_c = 0.4$ and $f_c = B \, t = |n_o - n_e| \, t = |1.54822 - 1.55746| \cdot 2.1 \, mm = 19.4 \, \mu m$ as the carrier frequency from Eq. (\ref{fc}).

The sinusoidal signal provides an interesting pattern to reconstruct, and its periodicity makes it easier for us to recognize how well a reconstruction reproduces the pattern.
Our simulations cover the band 400--800 nm, but the algorithms in this paper work over any arbitrary band; the mathematical derivations in Section \ref{SectionTheory} are not constrained by wavelength.
In this section, the ratio $f_{S0}/f_c = 0.4$ is chosen to be small enough so that Fourier reconstruction can produce a nearly ideal reconstruction if there is no noise.
As noise is added to the simulated measurement, the reconstruction deviates from ideal, and we wish to study the performance of Fourier reconstruction and CCSP as the signal-to-noise ratio degrades.

As described in Section \ref{sectionFR}, Fourier reconstruction applies filters to the interferogram for extracting channels. We specify these filters as
    \begin{equation}
        H_\text{LPF}(d) = \text{rect}\left(\frac{d}{\Delta}\right)
        \label{hlpf}
    \end{equation}
    \begin{equation}
        H_\text{BPF}(d) = \text{rect}\left(\frac{d - d_0}{\Delta}\right)
        \label{hbpf}
    \end{equation}
where $d_0 = 19.4 \, \mu m$ corresponds to the carrier frequency, $\Delta = 19.4 \, \mu m$ is width of the rectangle, $d$ is the optical path difference variable, and
    \begin{equation}
        \text{rect}(d) = \begin{cases}
            0, & \text{if } |d| \geq \frac{1}{2} \\
            1, & \text{if } |d| < \frac{1}{2}.
        \end{cases}
    \end{equation}
These filters are rectangular windows centered at $d_0$ with a width of $\Delta$.
The bandwidth of the filtered signal is $\Delta / 2$, and it is chosen to maintain equal spectral resolution in both channels as mentioned in Section \ref{sectionFR}.

 We set the threshold $\tau$ from Problem (\ref{MinimizationProblem}) as $\tau = 20 \, \mu m$, chosen to be slightly above the carrier frequency $f_c = 19.4 \, \mu m$.
 As noted in Section \ref{sectionCS}, setting the threshold close to the carrier frequency is one possible guideline.
 It helps to maximize the bandwidth of the Stokes parameters while mitigating any oscillatory artifacts in the reconstruction.
 Another possible way to tune the threshold is to set it to a high value and decrease it if high frequency oscillations are observed and not expected.

 Another reconstruction parameter is the number of Legendre polynomials, $L$.
 As described in Section \ref{sectionCS}, the Legendre polynomials help to model low order polynomial variations, including linear, quadratic, and cubic polynomials.
 Simulations can help to determine how many polynomials are needed for accurate reconstructions.
 As $L$ varies, we can observe how well the result matches the known input Stokes parameters.
 We find that $L = 5$ is sufficient to represent these lower order signals for our scenarios, and we use this value throughout the paper.

    The output of a channeled spectropolarimeter with noise can be modeled as
\begin{equation}
    y(\sigma) = \frac{1}{2} \left\{ S_0(\sigma) + S_1(\sigma) \text{cos}(2 \pi f_c \sigma) + S_2(\sigma) \text{sin}(2 \pi f_c \sigma) \right\} + n
    \label{eqn:ynoise}
\end{equation}
where we assume that the noise is an independent, identically distributed (IID) Gaussian random variable,
\begin{equation}
n \sim \mathcal N\left(\mu = 0, \sigma_n^2\right),
    \label{eqn:noise}
\end{equation}
with mean $\mu = 0$ and variance $\sigma_n^2$. In vector form, let $\lbf y \in \mathbb R^N$ denote the noisy output of the channeled spectropolarimeter with $i$th entry
\begin{equation}
    y_i = \frac{1}{2} \left\{ S_0(\sigma_i) + S_1(\sigma_i) \text{cos}(2 \pi f_c \sigma_i) + S_2(\sigma_i) \text{sin}(2 \pi f_c \sigma_i) \right\} + n_i
    \label{eqn:yinoise}
\end{equation}
where $\sigma_i$ is the $i$th wavenumber and $n_i$ is the $i$th noise sample for $i = 1, \dots, N$, with $N$ as the number of samples.
The measured output $\lbf y$ differs from the system model $\lbf y_{\text{model}}$ in Eq. (\ref{eqn:ymodeli}) by the added noise.

The first step in the simulation is to generate a measurement $y(\sigma)$ by plugging in the known Stokes parameters from Eqs. (\ref{eqn:s0snr})--(\ref{eqn:s2snr}), which we call \emph{ground truth}.
The next step is to create noise samples according to Eq. (\ref{eqn:noise}) over a variety of $\sigma_n$.
We try values of $\sigma_n$ ranging from 0 to 0.9 in increments of 0.05.
We estimate the noise power $P_n$ as
\begin{equation}
    P_n = \frac{1}{N} \sum_{i = 1}^{N} n_i^2.
    \label{eqn:pn}
\end{equation}
We estimate the signal power in a similar way:
\begin{equation}
    P_s = \frac{1}{N} \sum_{i = 1}^{N} y_i^2.
    \label{eqn:ps}
\end{equation}
We define the signal-to-noise ratio (SNR) with units of decibels (dB) as
\begin{equation}
    \text{SNR} = 10 \, \text{log}_{10} \frac{P_s - P_n}{P_n}.
    \label{eqn:snr}
\end{equation}

    We wish to compare Fourier reconstruction and CCSP using the simulated output of the channeled spectropolarimeter.
    Let $\lbf s_i^\text{GT}$, $\lbf s_i^\text{FR}$, and $\lbf s_i^\text{CCSP}$ denote the Stokes parameters corresponding to ground truth, Fourier reconstruction, and CCSP, respectively, for $i = 0, 1, 2$.
The metrics ``FR Fit'' and  ``CCSP Fit'' measure how closely Fourier reconstruction and CCSP match ground truth:
\begin{equation}
    \text{FR Fit}\left(\lbf s_i^\text{FR}, \lbf s_i^\text{GT}\right) = 1 - \frac{\left|\left|\lbf s_i^\text{FR} - \lbf s_i^\text{GT}\right|\right|}{\left|\left|\lbf s_i^{\text{GT}}\right|\right|}
    \label{eqn:FRFit}
\end{equation}
and
\begin{equation}
    \text{CCSP Fit}\left(\lbf s_i^\text{CCSP}, \lbf s_i^\text{GT}\right) = 1 - \frac{\left|\left|\lbf s_i^\text{CCSP} - \lbf s_i^\text{GT}\right|\right|}{\left|\left|\lbf s_i^{\text{GT}}\right|\right|}.
    \label{eqn:CSFit}
\end{equation}
For example, a fit of 100\% indicates that two waveforms perfectly match.
The percent change $\Delta_\%$ from ``FR Fit'' to ``CCSP Fit'' quantifies how much improvement CCSP provides:
\begin{equation}
    \Delta_\%(\text{FR Fit}, \text{CCSP Fit}) = \frac{\text{CCSP Fit} - \text{FR Fit}}{\text{FR Fit}} * 100.
    \label{eqn:Delta}
\end{equation}

\begin{figure}[!htb]
  \label{fig7}
  \begin{subfigure}[b]{0.5\linewidth}
    \centering
    \includegraphics[width=0.97\linewidth]{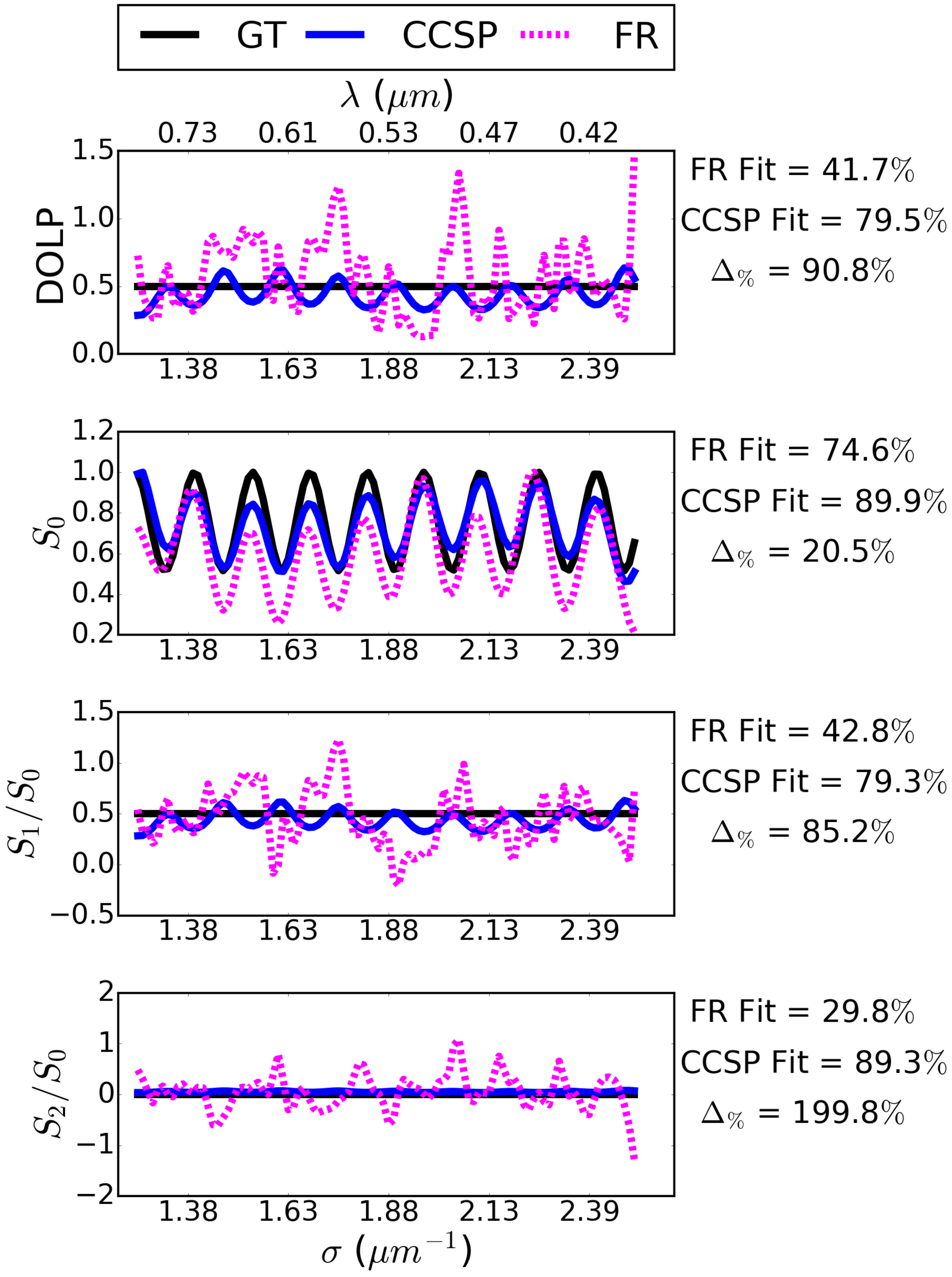}
    \caption{Reconstruction with SNR = 10.4 dB. This SNR is high enough that both algorithms clearly capture the sinusoidal signal.\\}
    \label{fig:snra}
  \end{subfigure}
  \begin{subfigure}[b]{0.5\linewidth}
    \centering
    \includegraphics[width=0.97\linewidth]{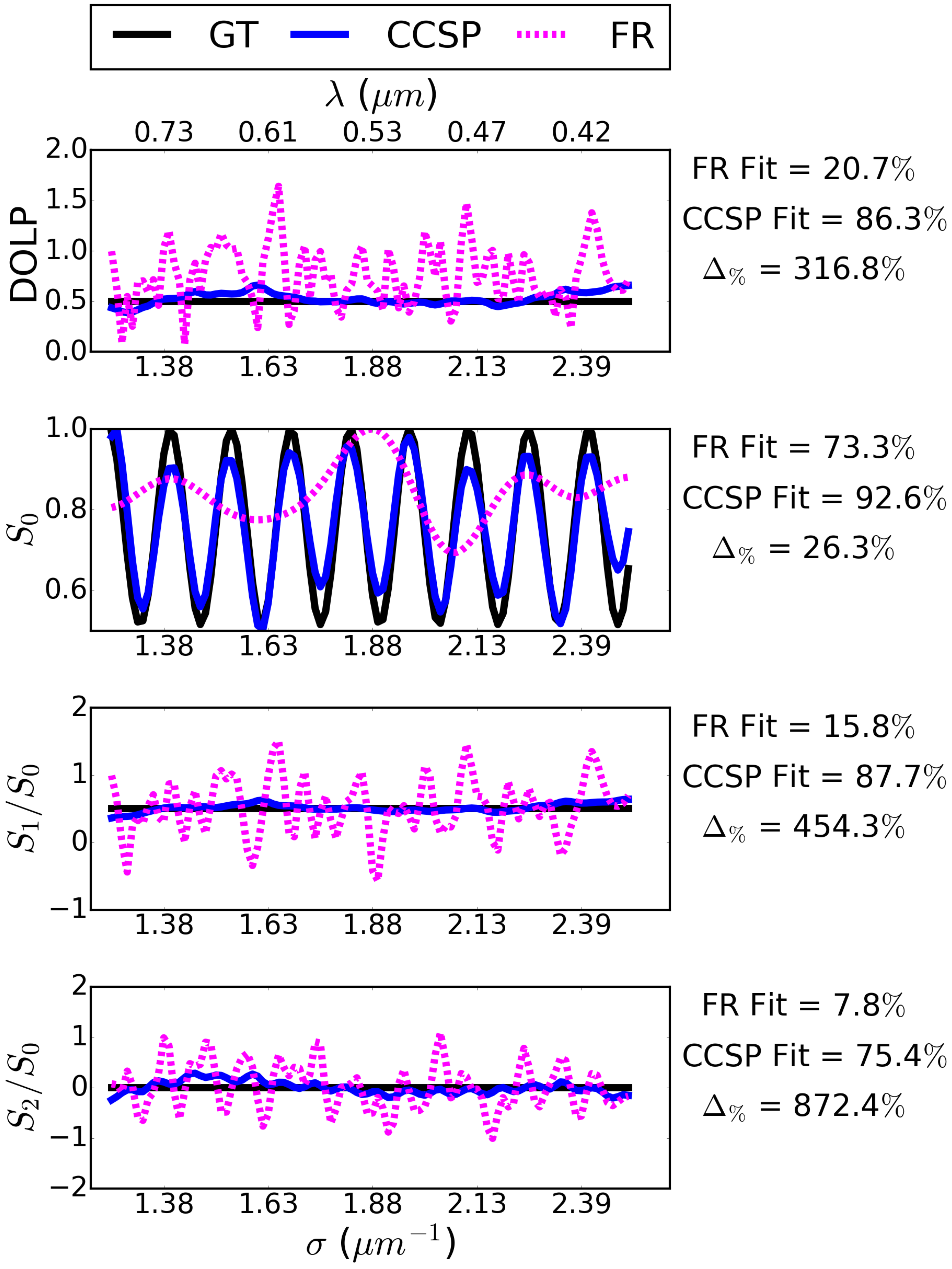}
    \caption{Reconstruction with SNR = 8.2 dB. This SNR is low enough that Fourier reconstruction fails to accurately capture the sinusoid, in contrast to the improved reconstruction of CCSP.}
    \label{fig:snrb}
  \end{subfigure}
      \caption{Reconstructions at two SNRs from a simulated measurement with added white Gaussian noise.
      The input Stokes parameters are defined in Eqs. (\ref{eqn:s0snr})--(\ref{eqn:s2snr}).
      The fit metrics for Fourier reconstruction and CCSP and the percent change $\Delta_\%$ from Fourier reconstruction to CCSP are defined in Eqs. (\ref{eqn:FRFit})--(\ref{eqn:Delta}).
      GT: Ground truth; FR: Fourier reconstruction; CCSP: Compressed channeled spectropolarimetry.}
      \label{fig:snr}
\end{figure}

\begin{figure}[!htb]
\centering
\includegraphics[width=0.4\linewidth]{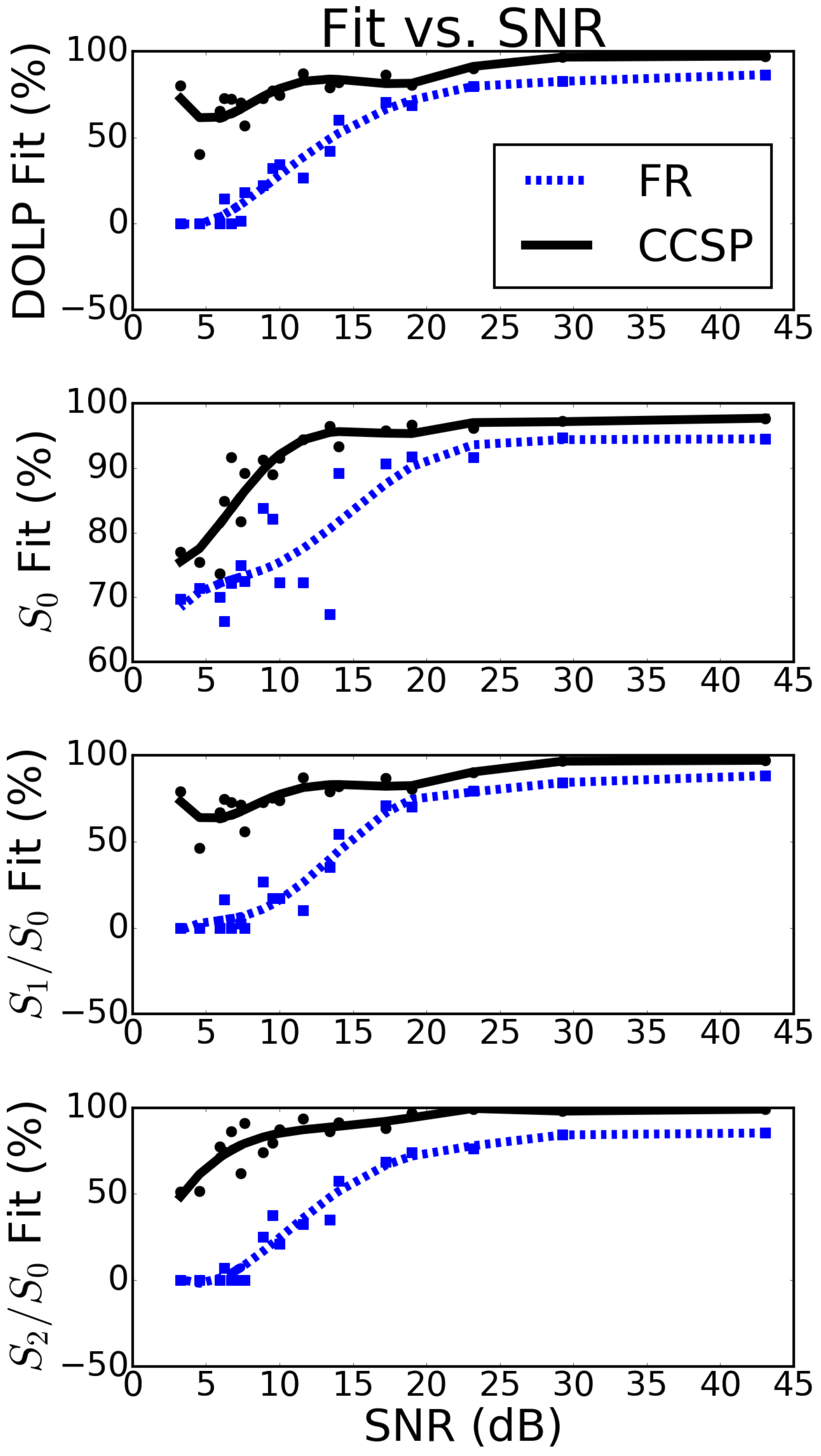}
\caption{Fit values at different SNRs, showing how well each reconstruction matches ground truth.
        We add noise to a simulated measurement, where the input Stokes parameters are defined in Eqs. (\ref{eqn:s0snr})--(\ref{eqn:s2snr}).
      As SNR decreases, CCSP outperforms Fourier reconstruction.
      The fit metrics for Fourier reconstruction and CCSP are defined in Eqs. (\ref{eqn:FRFit})--(\ref{eqn:Delta}).
      The lines show polynomial fits to the data points.
    FR: Fourier reconstruction; CCSP: Compressed channeled spectropolarimetry.}
    \label{fig:FitVsSNR}
\end{figure}

    Figure \ref{fig:snra} shows the reconstruction when SNR = 10.4 dB.
    We initialize the regularizer weight $\beta$ from Problem (\ref{MinimizationProblem}) as $\beta = 0.7$, and we will discuss guidelines for setting this parameter later.
    The output parameters include $S_0$, the normalized values $S_1 / S_0$ and $S_2 / S_0$, and the degree of linear polarization (DOLP), where
    \begin{equation}
        \text{DOLP}(\sigma) = \frac{\sqrt{S_1^2(\sigma) + S_2^2(\sigma)}}{S_0(\sigma)},
    \end{equation}
    and all values are plotted against wavenumber and wavelength.
    The ground truth values from Eqs. (\ref{eqn:s0snr})--(\ref{eqn:s2snr}) correspond to the black line.
    The results for Fourier reconstruction and CCSP are depicted as solid blue and dashed pink lines, respectively.
    On the right side of each plot, we calculate fit metrics according to Eqs. (\ref{eqn:FRFit})--(\ref{eqn:CSFit}) as percentages, and the percent change from Fourier reconstruction to CCSP from Eq. (\ref{eqn:Delta}).
    The fit metric for DOLP is calculated analogously to Eqs. (\ref{eqn:FRFit})--(\ref{eqn:CSFit}).
    CCSP better captures the $S_0$ cosine shape and shows less variation in $S_1 / S_0$, $S_2 / S_0$, and DOLP.

\begin{figure}[!htb]
\centering
\includegraphics[width=0.5\linewidth]{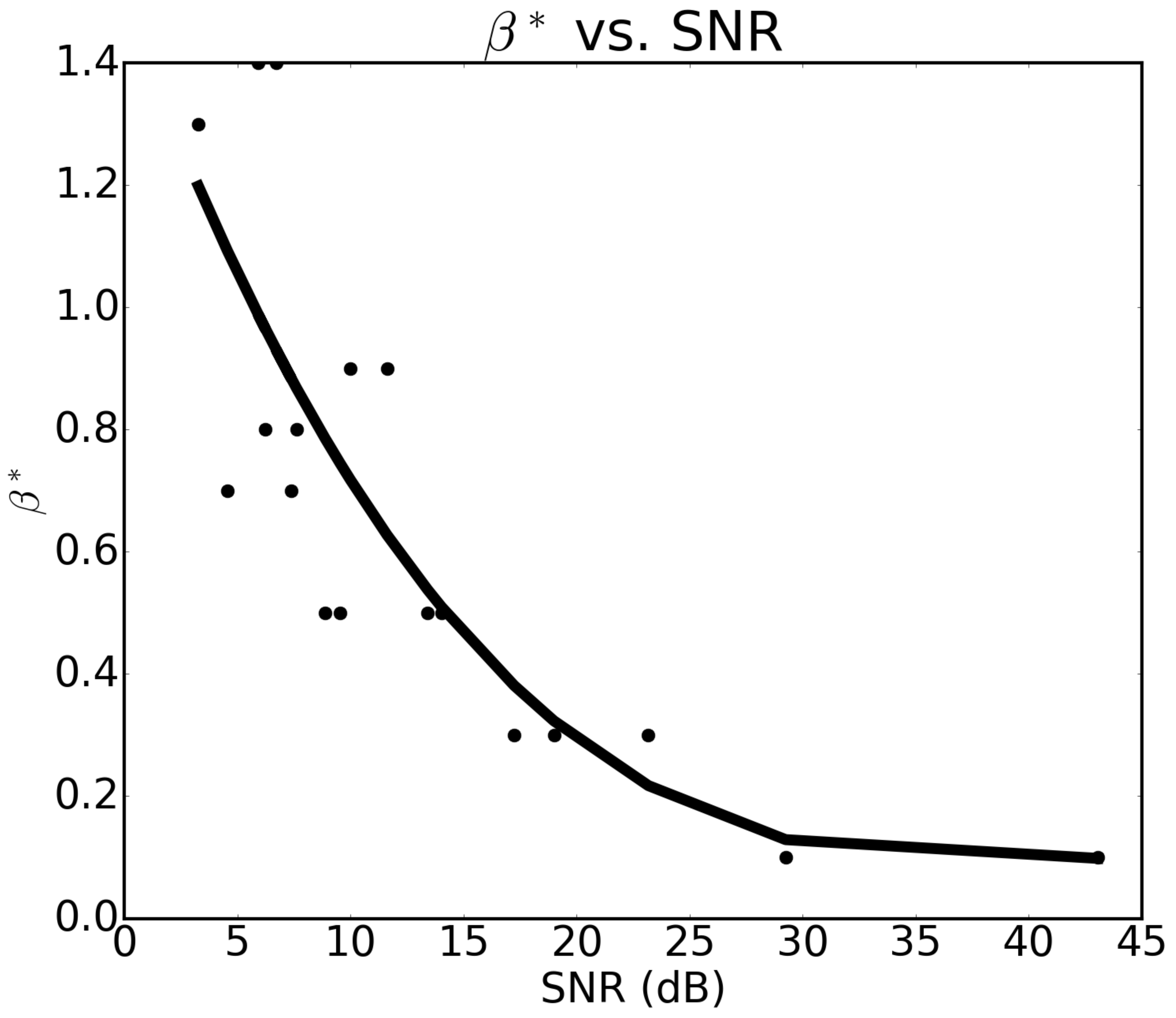}
    \caption{Optimal regularizer weights $\beta$ at different SNRs that resulted in the most accurate reconstructions.
    We add varying noise to a simulated measurement, where the input Stokes parameters are defined in Eqs. (\ref{eqn:s0snr})--(\ref{eqn:s2snr}).
    $\beta$ is the regularizer weight from Problem (\ref{MinimizationProblem}).
    At lower SNR, increasing the regularizer weight improves robustness to noise and results in more accurate reconstructions.
    The line is a polynomial fit to the data points.}
    \label{fig:BetaVsSNR}
\end{figure}

    Figure \ref{fig:snrb} shows the reconstruction when SNR = 8.3 dB.
    We initialize the regularizer weight $\beta$ from Problem (\ref{MinimizationProblem}) as $\beta = 0.8$, and we will discuss guidelines for setting this parameter later.
    As SNR decreases, the noise further contaminates the channels from Eqs. (\ref{eqn:frC0}) and (\ref{eqn:frC1}).
    As a result, the Fourier reconstruction fit values worsen compared to Fig. \ref{fig:snra}.
    The CCSP fit values are consistently better and show an even greater percent improvement in Fig. \ref{fig:snrb}.
    For example, CCSP captures the cosine shape in $S_0(\sigma)$, whereas the Fourier reconstruction result looks like a filtered and smoothed version.
    In this case, the noise degrades the peak of the sideband so that Fourier reconstruction algorithm detects an incorrect peak for filtering the channel.
    This example illustrates sensitivity to noise in Fourier reconstruction, whereas CCSP is more robust to low SNR.

    Figure \ref{fig:FitVsSNR} shows fits for different SNRs.
    Note that Fig. \ref{fig:snr} displays fits at two specific SNRs.
    In general, CCSP demonstrates better performance across all SNRs, and improvement even increases with higher noise levels.
    Note that this figure applies to the case study described in this section.
    Similar studies of SNR can be performed for other scenes using this example as a guide.

    As SNR changes, we wish to investigate which values of the regularizer weight $\beta$ result in the best performance.
    We test different SNRs by varying the standard deviation of the noise in Eq. (\ref{eqn:noise}).
    For each noise level, we try values of $\beta$ ranging from 0 to 1.4 in increments of 0.1.
    The best performing value of $\beta$, labeled as $\beta^\ast$, produces the reconstruction with the highest fit.
    Figure \ref{fig:BetaVsSNR} shows $\beta^\ast$ as a function of SNR.
    As SNR decreases, performance improves when $\beta$ increases, which corresponds to giving more weight to the regularization term.
    This simulation provides a guideline for choosing $\beta$ based on SNR; the values of $\beta$ in this paper are based on this guideline.
    This study provides an example of how to set up a simulation to test values of $\beta$, and similar studies can be applied to other scenes and noise models that may be of interest.

\subsection{Test cases with varying frequency}\label{SectionVaryFrequency}

\begin{figure}[!htb]
  \label{fig7}
  \begin{subfigure}[b]{0.5\linewidth}
    \centering
    \includegraphics[width=0.97\linewidth]{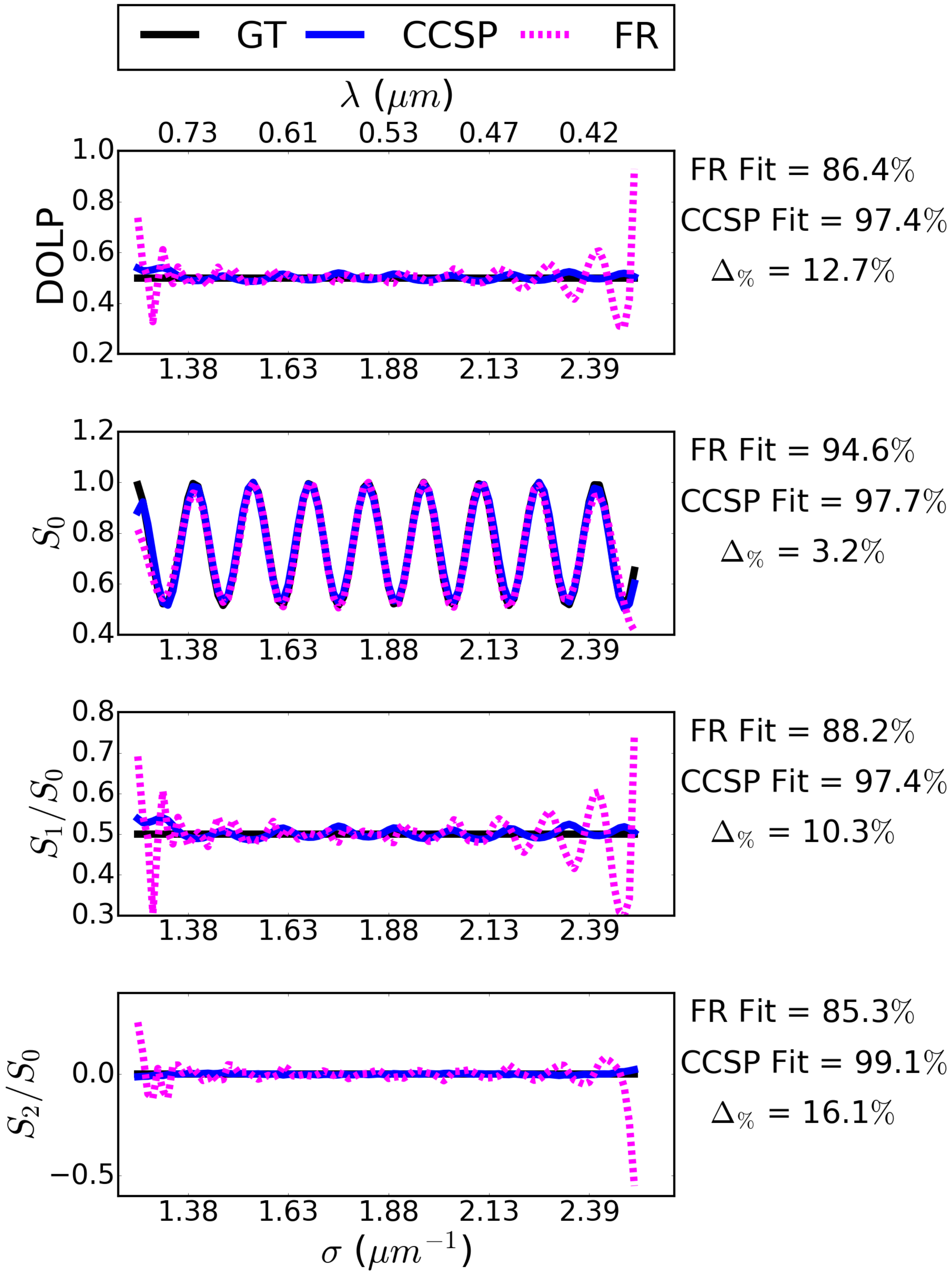}
      \caption{Reconstruction with $f_{S0} / f_c = 0.4$.
      The frequency of $S_0(\sigma)$ is low enough so that both algorithms capture the sinusoid.
      Fourier reconstruction exhibits edge artifacts where the signal oscillates because the Fourier transform imposes an assumption of periodic boundary conditions, whereas CCSP avoids these nonphysical oscillations.}
    \label{fig:freq0400}
  \end{subfigure}
  \begin{subfigure}[b]{0.5\linewidth}
    \centering
    \includegraphics[width=0.97\linewidth]{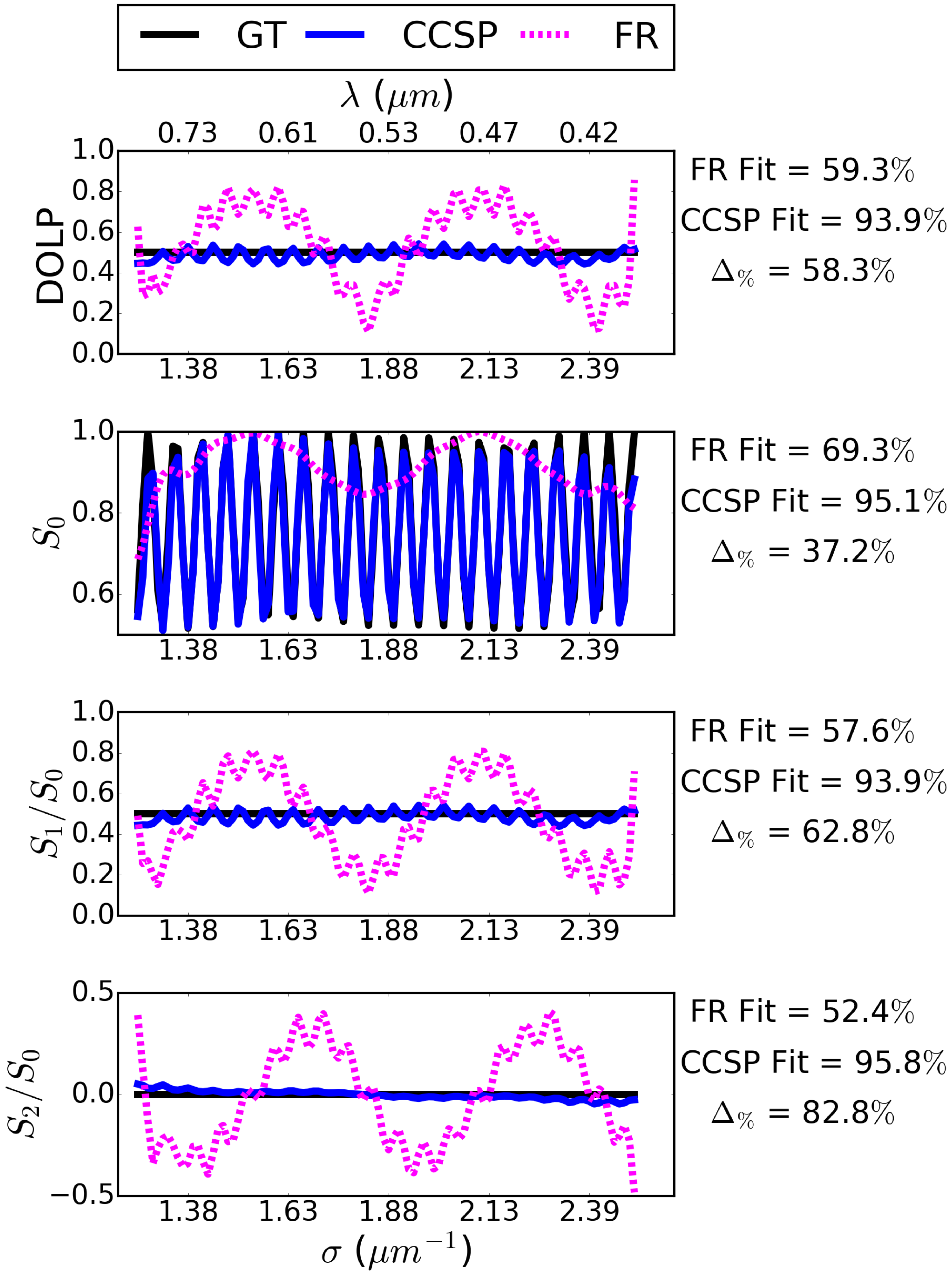}
      \caption{Reconstruction with $f_{S0} / f_c = 0.9$.
      The frequency of $S_0(\sigma)$ is too high to be reconstructed accurately by Fourier reconstruction because it falls outside of the Fourier domain window, whereas CCSP does not impose a windowing constraint.\\\\}
    \label{fig:freq0900}
  \end{subfigure}
      \caption{Reconstructions at two simulated values of $f_{S0} / f_c$.
      The frequency of $S_0(\sigma)$ varies relative to the carrier frequency of the measurement.
      The input Stokes parameters are defined in Eqs. (\ref{eqn:s0snr})--(\ref{eqn:s2snr}).
      The fit metrics for Fourier reconstruction and CCSP and the percent change $\Delta_\%$ from Fourier reconstruction to CCSP are defined in Eqs. (\ref{eqn:FRFit})--(\ref{eqn:Delta}).
      GT: Ground truth; FR: Fourier reconstruction; CCSP: Compressed channeled spectropolarimetry.}
  \end{figure}

    Our goal in this section is to reconstruct Stokes parameters that have varying frequency.
    The ground truth values of $S_0(\sigma)$, $S_1(\sigma)$, and $S_2(\sigma)$ have the same form as Eqs. (\ref{eqn:s0snr})--(\ref{eqn:s2snr}), and the values of $a_1$, $a_2$, $b$, and $f_c$ are the same.
    To generate the sample measurement, we set the noise to zero ($n = 0$ in Eq. (\ref{eqn:ynoise})) in order to focus the study on varying frequency.
    For the reference measurement, we set $S_0^R(\sigma) = 1$, $S_1^R(\sigma) = 1$, and $S_2^R(\sigma) = 0$, and the phase of the optical system can be estimated according to Eq. (\ref{eqn:phihat}).
    We vary the ratio of $f_{S0} / f_c$ from 0 to 1.2 in increments of 0.1.
    For each ratio, we measure the fit of the reconstruction in order to compare how well Fourier reconstruction and CCSP match the ground truth.
    Note that we use the same filter windows $H_\text{LPF}(d)$ and $H_\text{BPF}(d)$ for Fourier reconstruction as in Section \ref{SectionSNR}.

    We will specify the regularizer weight $\beta$ and the threshold $\tau$.
    This simulation has high SNR since it is noiseless, so we set $\beta = 10$ using Section \ref{SectionSNR} as a guide.
    Since frequency is varying in this simulation, we choose to remove the constraint in Problem (\ref{MinimizationProblem}), so $\tau$ is no longer a parameter.
    By removing the constraint, we do not need to set $\tau$ depending on the ratio of $f_{S0} / f_c$.
    For any algorithm, it is desirable to set as few parameters as possible, and the parameters should work over a broad range of cases.
    We will show that these parameter settings will work in all the test cases in this simulation.

\begin{figure}[!htb]
    \centering
    \includegraphics[width=0.4\linewidth]{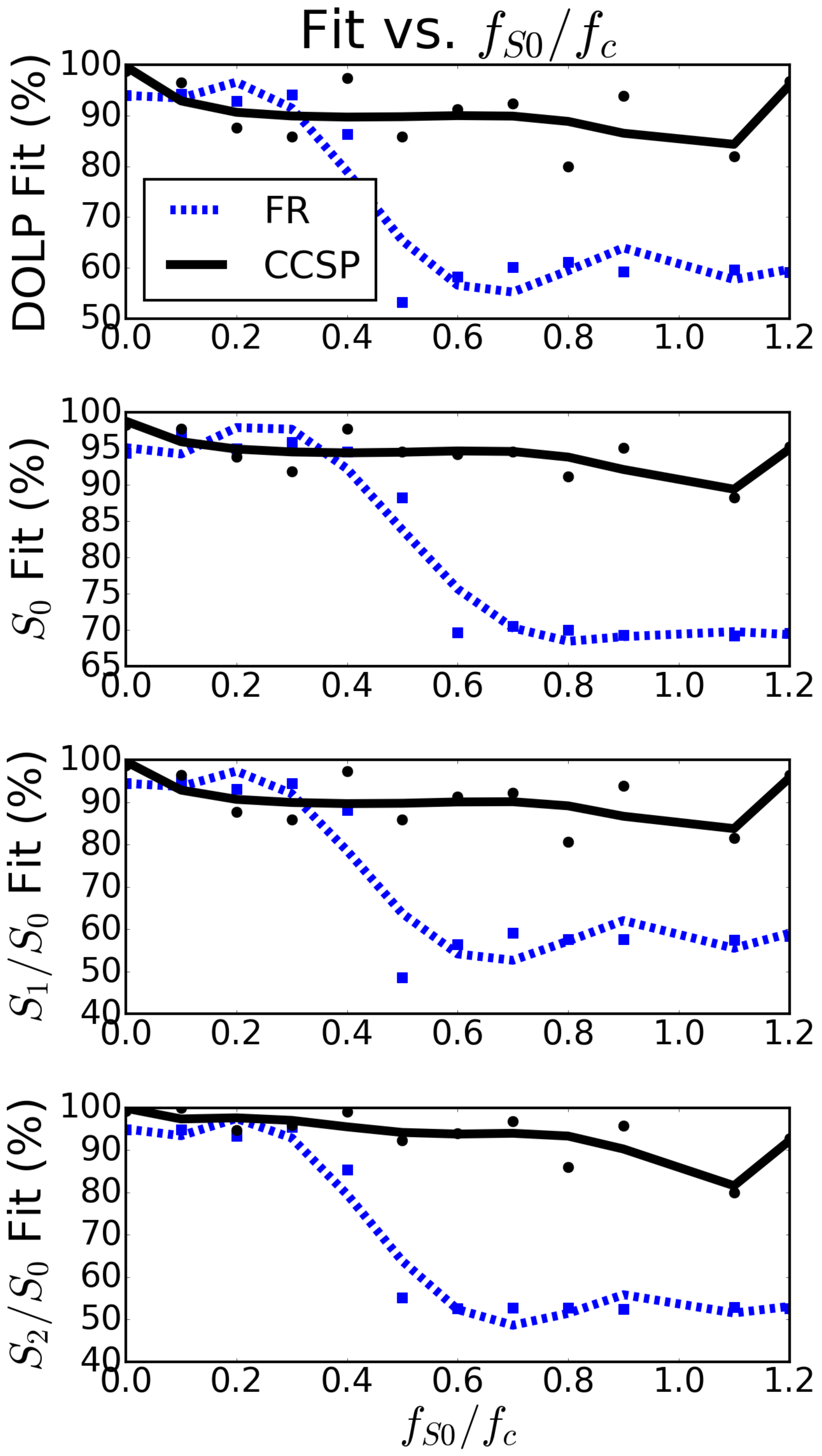}
    \caption{Fit at different values of $f_{S0} / f_c$, showing how well each reconstruction matches ground truth.
      The frequency of $S_0(\sigma)$ varies relative to the carrier frequency of the measurement.
      The input Stokes parameters are defined in Eqs. (\ref{eqn:s0snr})--(\ref{eqn:s2snr}).
      As $f_{S0}$ increases, the Fourier reconstruction degrades because it imposes a windowing constraint that cuts off high frequency details, whereas CCSP does not impose this constraint and can recover higher frequencies.
      The fit metrics for Fourier reconstruction and CCSP and the percent change $\Delta_\%$ from Fourier reconstruction to CCSP are defined in Eqs. (\ref{eqn:FRFit})--(\ref{eqn:Delta}).
      FR: Fourier reconstruction; CCSP: Compressed channeled spectropolarimetry.}
    \label{fig:FitVsFreq}
\end{figure}

    Figure \ref{fig:freq0400} shows the reconstruction when $f_{S0} / f_c = 0.4$.
    As in Section \ref{SectionSNR}, we calculate fits for each parameter using Eqs. (\ref{eqn:CSFit})--(\ref{eqn:FRFit}) and the percent improvement from Fourier reconstruction to CCSP according to Eq. (\ref{eqn:Delta}).
    Both Fourier reconstruction and CCSP capture the cosine shape of $S_0$ well.
    However, Fourier reconstruction has reconstruction artifacts near the ends of the spectra in DOLP, $S_1 / S_0$, and $S_2 / S_0$.
    For example, the Fourier reconstruction values in $S_1 / S_0$ grow in oscillations near the edges.
    These effects are due to properties of the Fourier transform, which assumes periodic boundary conditions, and they highlight a limitation of Fourier reconstruction.
    In contrast, CCSP is not limited by any assumptions on periodicity.
    As a result, CCSP is more accurate, as reflected in the percent improvement over Fourier reconstruction.

    Figure \ref{fig:freq0900} shows the reconstruction when $f_{S0} / f_c = 0.9$.
    Fourier reconstruction fails to capture the full cosine modulation in $S_0$, which occurs because the $S_0$ frequency falls outside of the channel.
    This effect highlights another drawback of Fourier reconstruction: the results are highly dependent on the window function, described in Section \ref{sectionFR}.
    In contrast, CCSP does not require the choice of a window function, and it is able to improve reconstruction accuracy.

    Figure \ref{fig:FitVsFreq} shows the calculated fit at different values of $f_{S0} / f_c$.
    Note Figs. \ref{fig:freq0400} and \ref{fig:freq0900} display reconstructions at two specific values of $f_{S0} / f_c$.
    At lower frequencies of $S_0(\sigma)$, both reconstructions perform well.
    For example, at $f_{S0} / f_c = 0.3$, Fourier reconstruction fits ground truth slightly better in $S_0(\sigma)$ than CCSP, but both fit values are above 90\%.
    The general trend is that Fourier reconstruction degrades when the ratio $f_{S0} / f_c$ increases beyond 0.5, and this value corresponds to the channel boundary.
    As described in Section \ref{sectionFR}, Fourier reconstruction chooses the peak of the sideband as the center of the channel $C_1$, and the filter window is commonly chosen to be the same length as the window for $C_0$ to maintain equal spectral resolution in both channels.
    Even if the width of the window for $C_0$ were increased to accomodate $f_{S0}$, it would cut off frequency content for $C_1$.
    In contrast, CCSP does not require this design choice, so it is able to maintain high accuracy across various values of $f_{S0} / f_c$.

\section{Experiment}\label{SectionExperiment}

In this section we will present experimental measurements from a channeled spectropolarimeter using a variety of samples.
The samples under test are not temporally dynamic.
To quantify the performance increase of our algorithm on the hardware implementation, we would like to compare against known, ground truth Stokes parameters.
We have built a rotating polarizer spectropolarimeter to estimate the ground truth.
Since this instrument generates channels in the temporal domain, we assume that there are no estimation artifacts with stationary samples.
We will show that CCSP mitigates many artifacts seen in Fourier reconstruction, including signal falloff at spectral edges, noise sensitivity, nonphysical oscillations in birefringent samples, and bandwidth limitations caused by windowing.

\begin{figure}[!htb]
    \centering
    \includegraphics[width=0.9\linewidth]{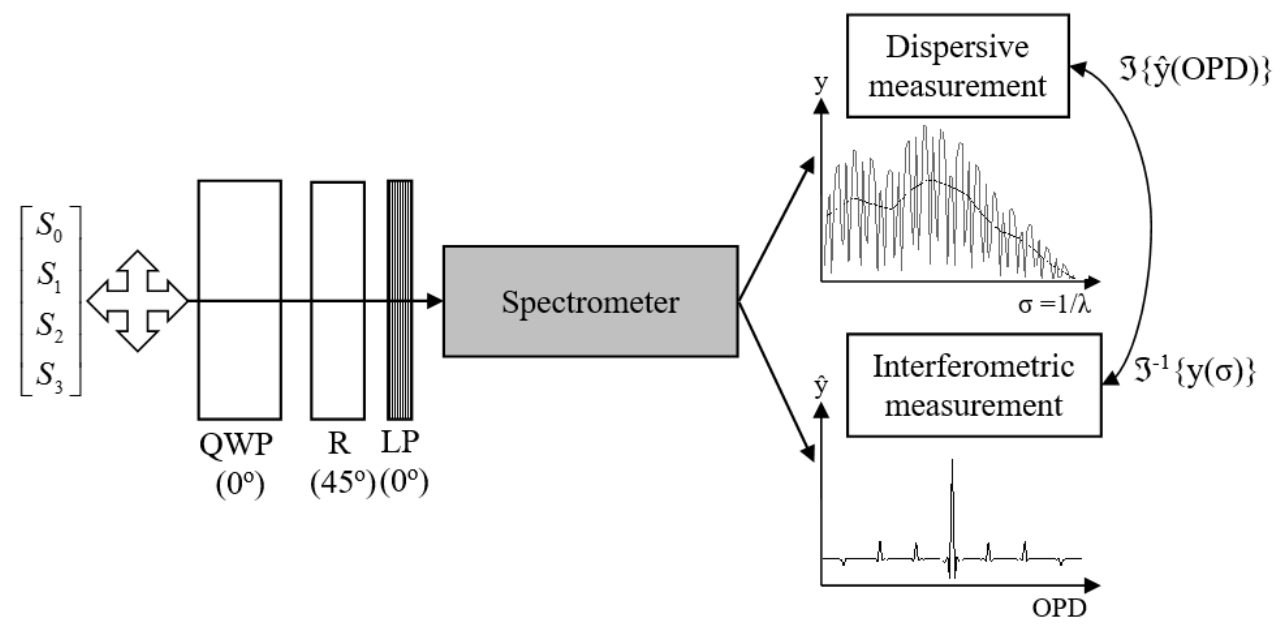}
    \caption{Experimental setup of the channeled spectropolarimeter. OPD: Optical path difference. QWP: Quarter wave plate; R: Retarder; LP: Linear polarizer.}
    \label{fig:setup}
\end{figure}

\begin{center}
    \begin{table}[!htb]
        \begin{tabular}{ccc}
    \toprule
    Sample Number & Sample & Measured polarizer angles ($^\circ$) \\
    \midrule
    1 & LP & $0^\circ$--$180^\circ$ in $22.5^\circ$ increments\\
    \midrule
    2 & Spectral Filter - LP & $0^\circ$--$180^\circ$ in $22.5^\circ$ increments\\
    \midrule
    3 & Spectral Filter - LP - Retarder & $0^\circ$--$180^\circ$ in $22.5^\circ$ increments \\
    \midrule
    4 & LP - Retarder & $0^\circ$--$180^\circ$ in $22.5^\circ$ increments \\
    \bottomrule
        \end{tabular}
        \caption{Samples used for experimental measurements. Each sample has a linear polarizer, and we rotate this polarizer from $0^\circ$ to $180^\circ$ in $22.5^\circ$ increments. We measure a total of 36 test cases, corresponding to the 4 samples and 9 polarizer angles per sample. LP: Linear polarizer.}
    \label{SampleTable}
    \end{table}
\end{center}

Figure \ref{fig:setup} shows the general setup.
Our light source is a QTH lamp.
The optical system consists of an achromatic quarter wave plate (QWP), retarder (R), and linear polarizer (LP), followed by a spectrometer (Ocean Optics HR4000).
We use a quartz retarder with thickness $t = 3.40$ mm and anti-reflective coating for 400--1000 nm.
The orientations of the fast axes of the quarter wave plate and retarder are 0$^\circ$ and 45$^\circ$.
All angles are relative to the transmission axis of the polarizer.

\begin{figure}[!tbh]
  \begin{subfigure}[b]{0.5\linewidth}
    \centering
    \includegraphics[width=0.97\linewidth]{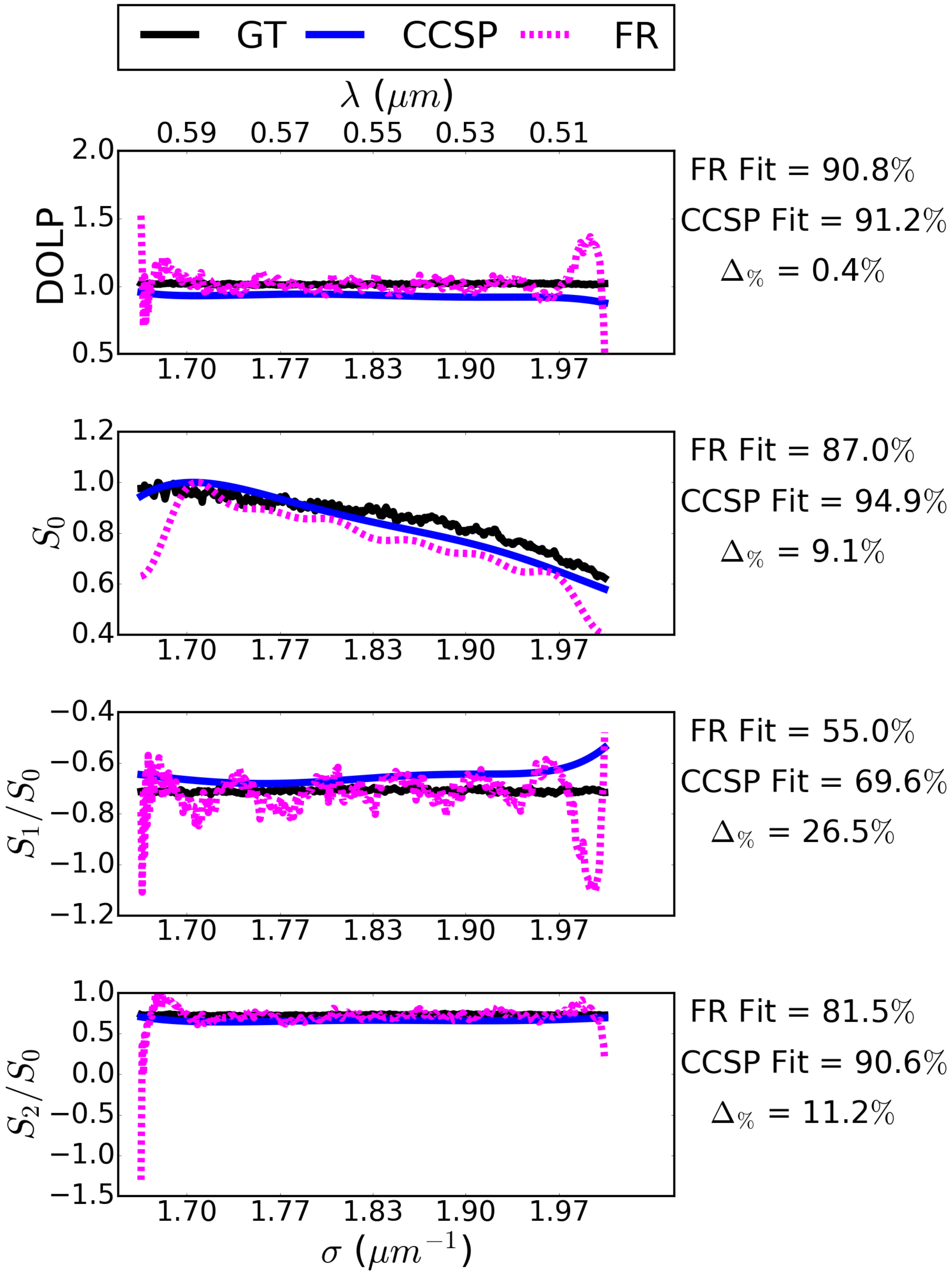}
      \caption{Reconstruction of sample 1 (LP). CCSP produces a more accurate, smoother reconstruction. Note $S_0(\sigma)$ shows the spectral shape of the QTH light source, and the DOLP is uniform because of the linear polarizer.}
    \label{fig:stokes1}
    \vspace{4ex}
  \end{subfigure}
  \begin{subfigure}[b]{0.5\linewidth}
    \centering
    \includegraphics[width=0.97\linewidth]{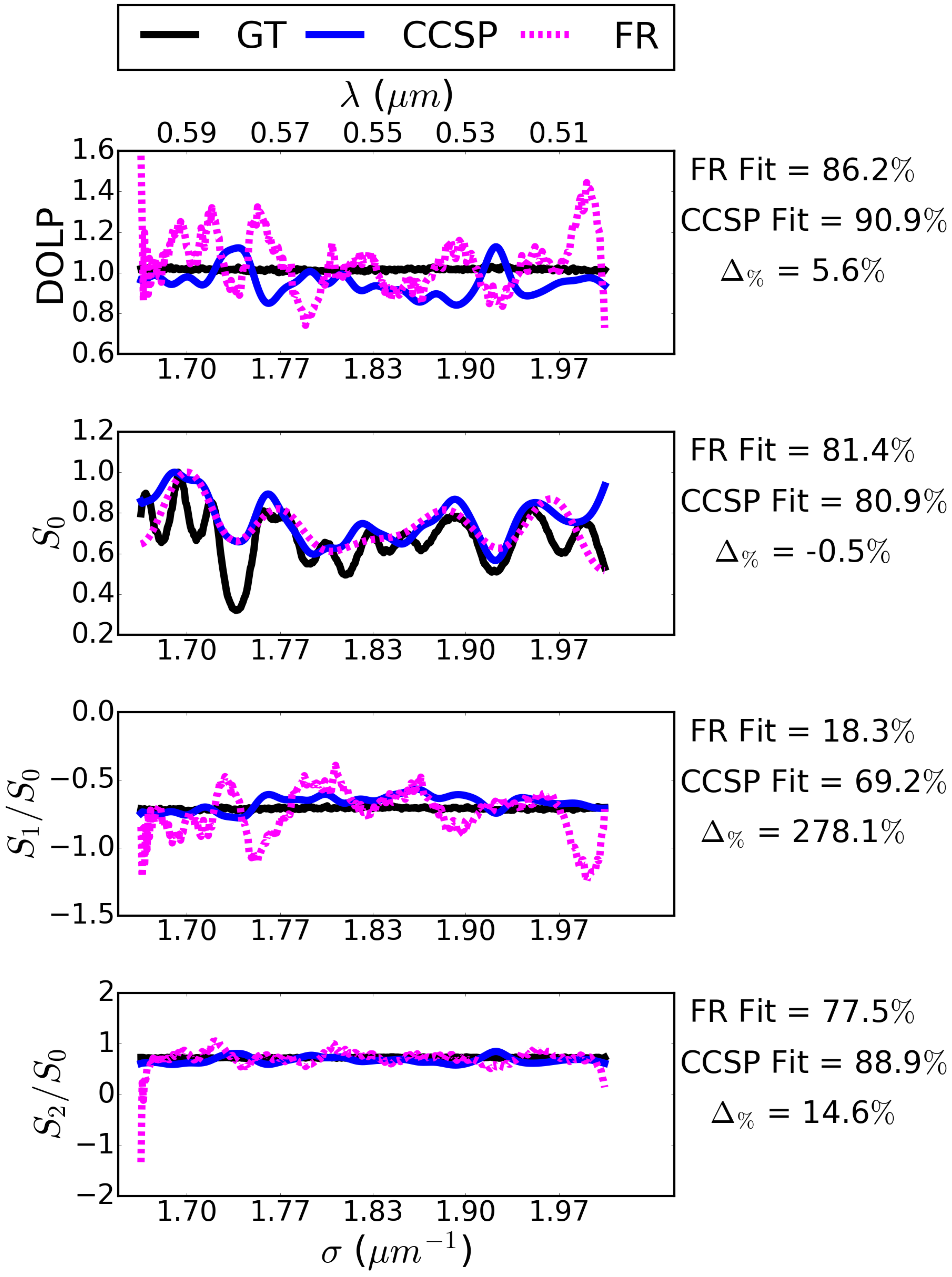}
      \caption{Reconstruction of sample 2 (Filter - LP). CCSP produces a more accurate, smoother reconstruction. Note $S_0(\sigma)$ shows the shape of the spectral filter, and the DOLP is uniform because of the linear polarizer.}
    \label{fig:stokes2}
    \vspace{4ex}
  \end{subfigure}
\end{figure}
\begin{figure}[!bth]\ContinuedFloat
  \begin{subfigure}[b]{0.5\linewidth}
    \centering
    \includegraphics[width=0.97\linewidth]{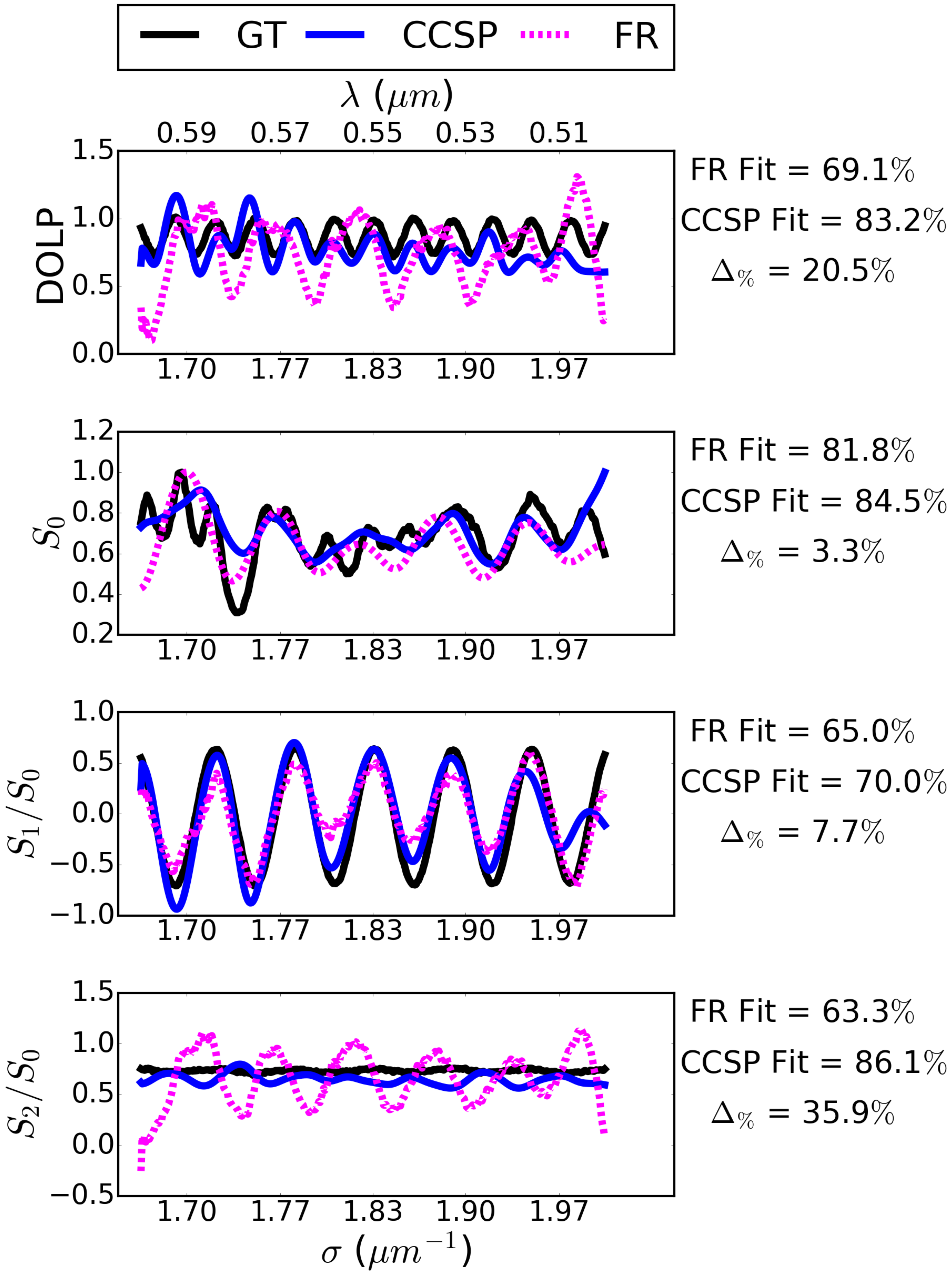}
    \caption{Reconstruction of sample 3 (Filter - LP - R). The Fourier reconstruction exhibits nonphysical oscillations in $S_2 / S_0$. $S_0(\sigma)$ shows the shape of the spectral filter, and DOLP varies sinusoidally due to the retarder.}
    \label{fig:stokes3}
  \end{subfigure}
  \begin{subfigure}[b]{0.5\linewidth}
    \centering
    \includegraphics[width=0.97\linewidth]{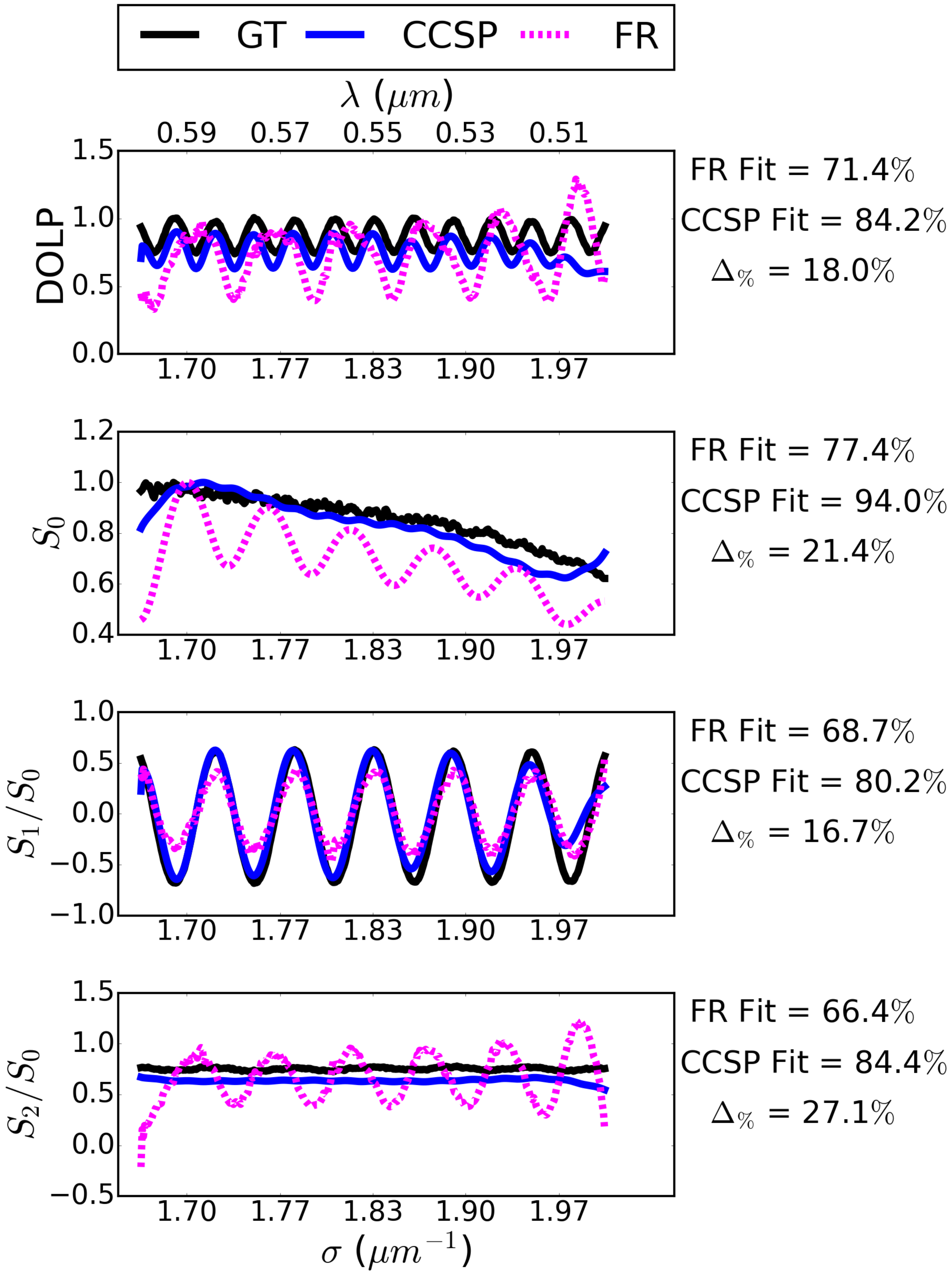}
      \caption{Reconstruction of sample 4 (LP - R).
      The Fourier reconstruction exhibits nonphysical oscillations in $S_0$ and $S_2 / S_0$.
      $S_0(\sigma)$ shows the shape of the QTH light source, and DOLP varies sinusoidally due to the retarder.}
    \label{fig:stokes4}
  \end{subfigure}
    \caption{Reconstructions of four measured samples with the sample polarizer oriented at $67.5^\circ$.
           We compare ground truth from a rotating polarizer spectropolarimeter with reconstructions from a channeled spectropolarimeter.
        The fit metrics for Fourier reconstruction and CCSP and the percent change $\Delta_\%$ from Fourier reconstruction to CCSP are defined in Eqs. (\ref{eqn:FRFit})--(\ref{eqn:Delta}).
      The samples are summarized in Table \ref{SampleTable}.
      GT: Ground truth; FR: Fourier reconstruction; CCSP: Compressed channeled spectropolarimetry.}
  \label{fig:stokes}
\end{figure}

    \begin{figure}[!htb]
  \begin{subfigure}[b]{0.5\linewidth}
    \centering
    \includegraphics[width=0.97\linewidth]{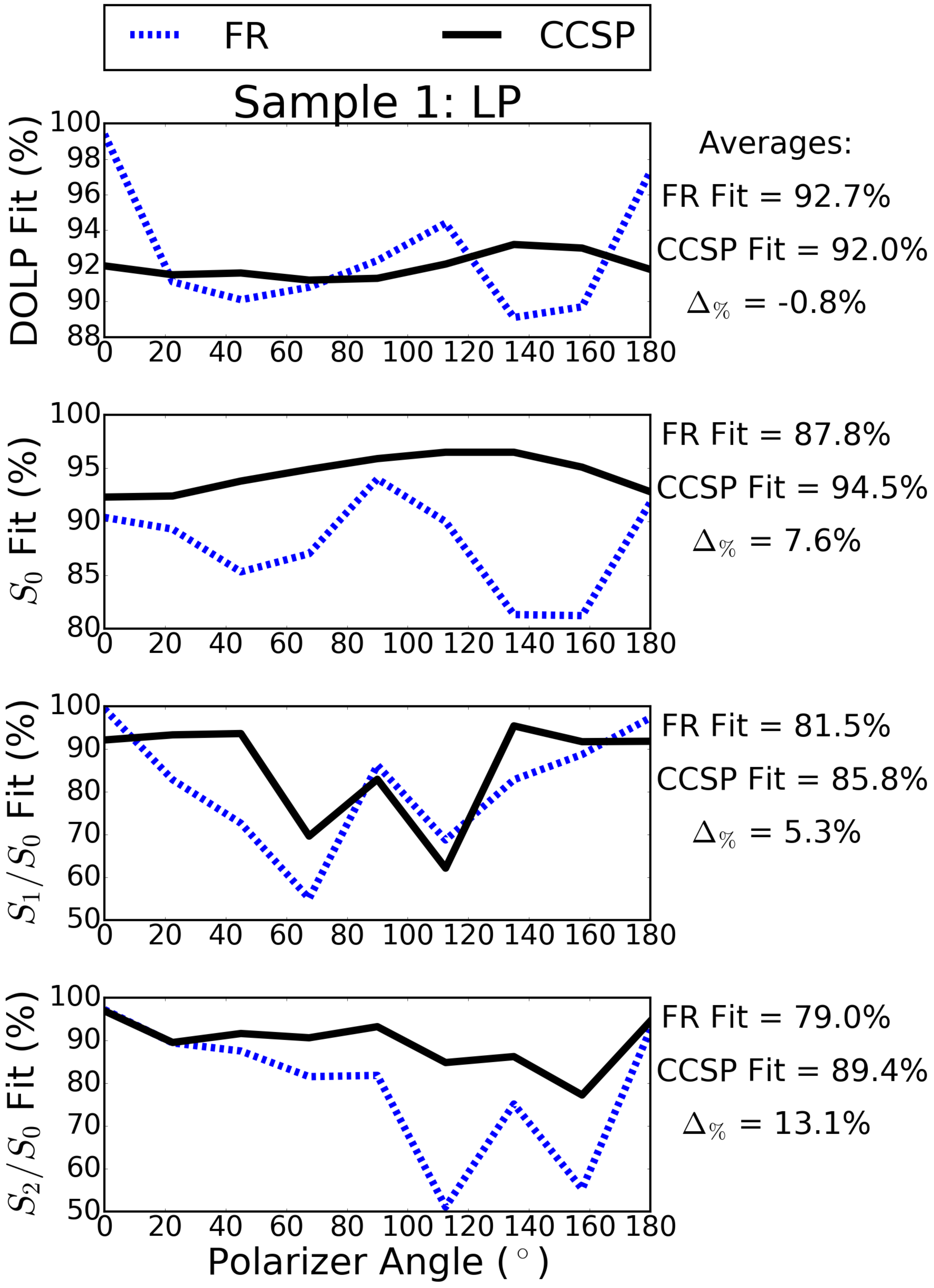}
      \caption{Fit vs. angle for sample 1 (LP)}
    \label{fig:fitangles1}
    \vspace{4ex}
  \end{subfigure}
  \begin{subfigure}[b]{0.5\linewidth}
    \centering
    \includegraphics[width=0.97\linewidth]{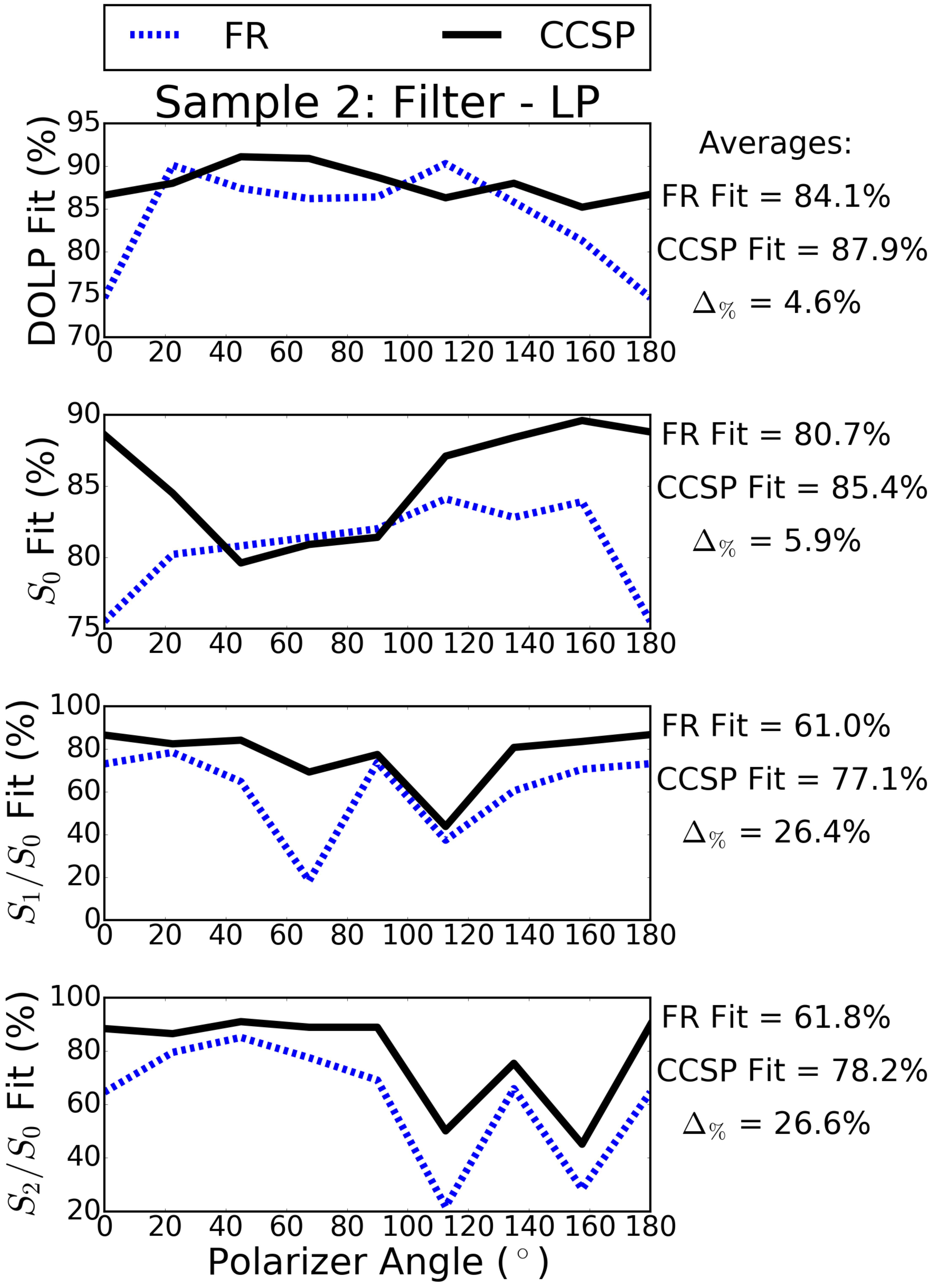}
    \caption{Fit vs. angle for sample 2 (Filter - LP)}
    \label{fig:fitangles2}
    \vspace{4ex}
  \end{subfigure}
\end{figure}
\begin{figure}[!htb]\ContinuedFloat
  \begin{subfigure}[b]{0.5\linewidth}
    \centering
    \includegraphics[width=0.97\linewidth]{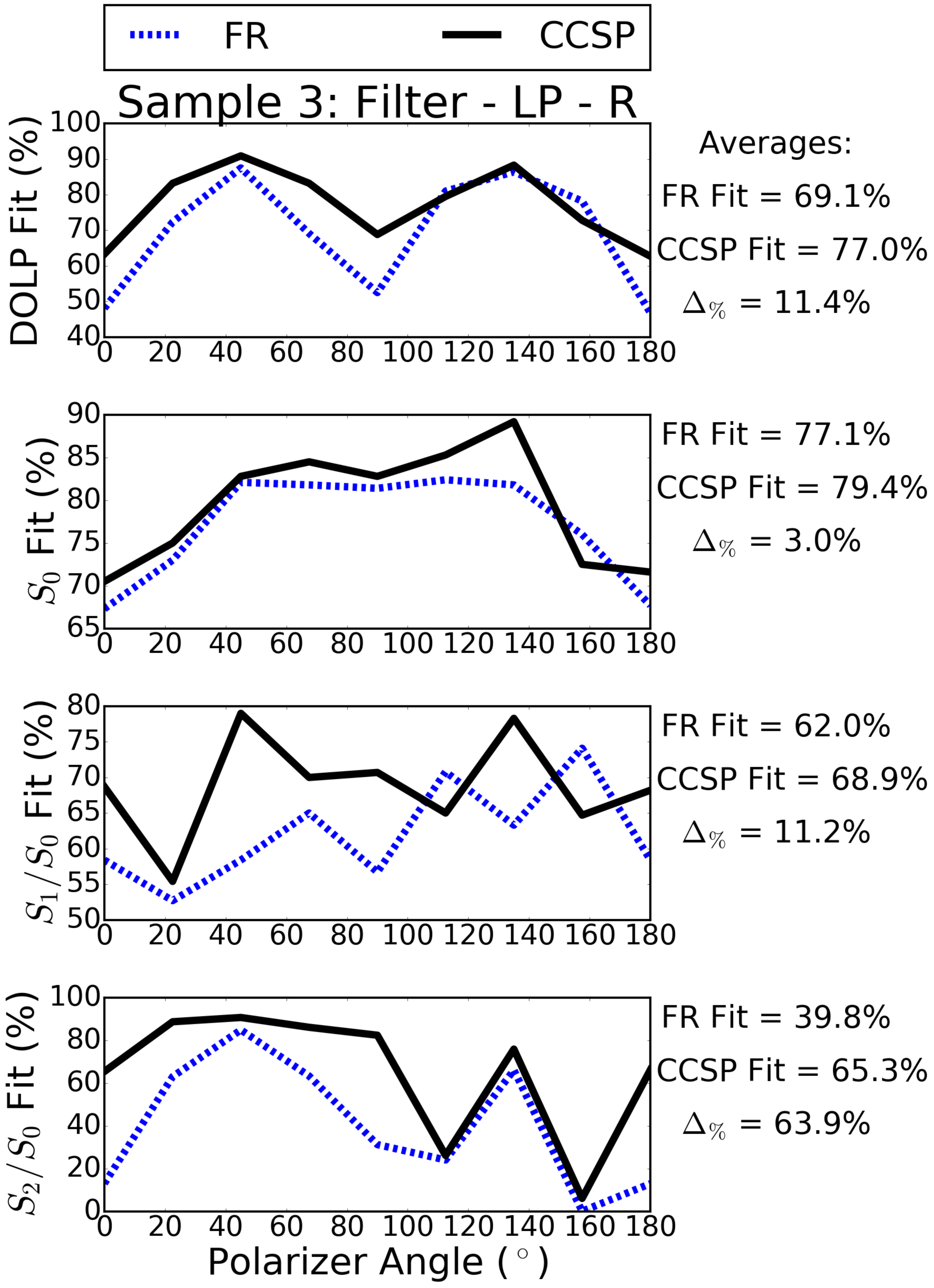}
    \caption{Fit vs. angle for sample 3 (Filter - LP - R)}
    \label{fig:fitangles3}
  \end{subfigure}
  \begin{subfigure}[b]{0.5\linewidth}
    \centering
    \includegraphics[width=0.97\linewidth]{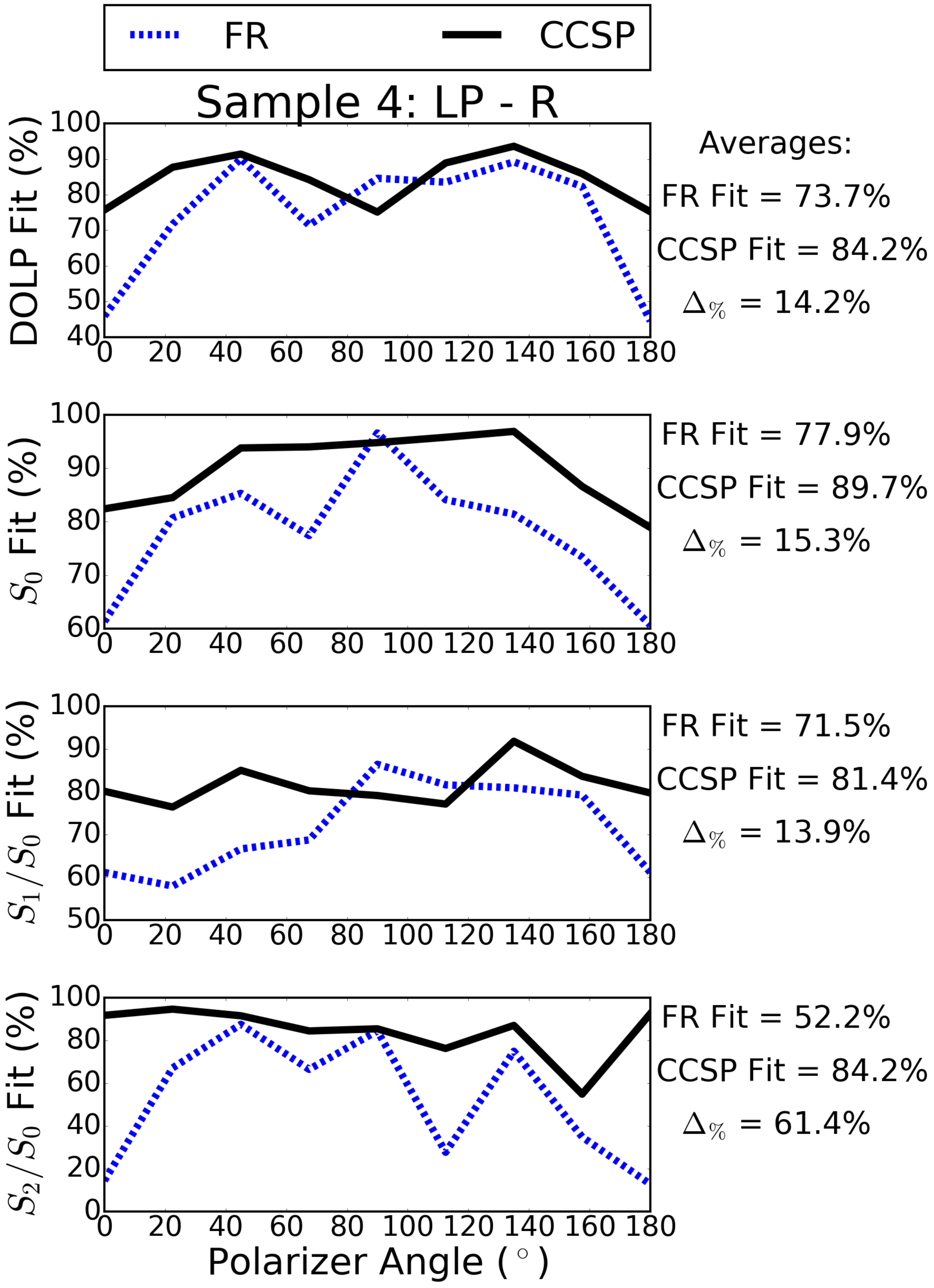}
    \caption{Fit vs. angle for sample 4 (LP - R)}
    \label{fig:figangles4}
  \end{subfigure}
  \caption{Fit values for four measured samples over varying sample polarizer angles, showing how well the reconstructions match ground truth.
        The fit metrics for Fourier reconstruction and CCSP are defined in Eqs. (\ref{eqn:FRFit})--(\ref{eqn:Delta}).
        The average fits over all angles are shown on the righthand side of the plots for each reconstruction.
        Here $\Delta_\%$ denotes the percent change of the averaged fits from Fourier reconstruction to CCSP.
      Note that Fig. \ref{fig:stokes} shows reconstructions with the sample polarizer angles oriented at $67.5^\circ$, and this figure plots the fit metrics over all measured polarizer angles.
      The samples are summarized in Table \ref{SampleTable}.
    FR: Fourier reconstruction; CCSP: Compressed channeled spectropolarimetry.}
  \label{fig:fitangles}
\end{figure}

\pgfplotstableset{
    highlightrow/.style={
        postproc cell content/.append code={
           \count0=\pgfplotstablerow
            \advance\count0 by1
            \ifnum\count0=#1
            \pgfkeysalso{@cell content=\textbf{##1}}%
            \fi
        },
    },
}

\begin{center}
    \begin{table}[!htb]
\pgfplotstabletypeset[
    highlightrow={3},
    highlightrow={6},
    highlightrow={9},
    highlightrow={12},
    column type=,
begin table={\begin{tabularx}{\textwidth}{bssss}},
end table={\end{tabularx}},
string type,
    columns/c1/.style={column name=(1) LP},
    columns/c2/.style={column name=(2) F},
    columns/c3/.style={column name=(3) F},
    columns/c4/.style={column name=(4) F},
every head row/.style={before row={\toprule},after row=\midrule},
every last row/.style={after row={\toprule}},
    every nth row={3}{before row=\midrule},
]\avgtable
    \caption{Fit values for each sample, averaged over all sample polarizer angles.
    The fit values quantify how well each reconstruction matches ground truth.
    This table summarizes the fit values that appear in Fig. \ref{fig:fitangles}.
    All quantites are percentages.
    The columns indicate which sample is measured.
    Each row shows statistics on reconstruction fit.
    FR: Fourier reconstruction; CCSP: Compressed channeled spectropolarimetry.}
    \label{StatisticsTable}
\end{table}
\end{center}

We measure a variety of samples to test the robustness of the reconstruction algorithms.
The first sample is a linear polarizer, which we expect to have a uniform DOLP.
The second sample is a spectral filter followed by a polarizer, which has a uniform DOLP and a modulated spectral shape based on the characteristics of the spectral filter.
The spectral filter is a long pass dichroic mirror (DMLP1180, Thorlabs) used in the transition band of the filter for its spectral shape.
The third sample is a spectral filter, polarizer, and retarder, which has a sinusoidal DOLP due to the retarder and a modulated spectral shape from the filter.
The fourth sample is a polarizer followed by a retarder, which has a sinusoidal DOLP and smooth spectral shape.
For the reference measurement, the channeled spectropolarimeter measures a horizontal polarizer, and we produce an estimated phase of the optical system, $\widehat{\phi}(\sigma)$, according to Eq. (\ref{eqn:phihat}).

Each of the four samples has a polarizer, and we rotate this polarizer at angles ranging from $0^\circ$ to $180^\circ$ in $22.5^\circ$ increments, for nine angles per sample.
Therefore, there are 36 total test cases.
Each \emph{test case} corresponds to a particular combination of a sample and polarizer angle.
Table \ref{SampleTable} summarizes the samples and measured polarizer angles.

    For each test case, we take measurements from two instruments: the rotating polarizer spectropolarimeter and the channeled spectropolarimeter.
    For the rotating polarizer spectropolarimeter described in Section \ref{SectionRotatingPolarizer}, we rotate the polarizer through $A = 9$ angles ranging from $0^\circ$ to $160^\circ$ in $20^\circ$ increments.
    The reconstruction from these measurements serves as ground truth.
    For the channeled spectropolarimeter, we wish to compare two methods for reconstructing Stokes parameters: Fourier reconstruction and CCSP.
    We evaluate the quality of the reconstructions by how well they match ground truth.

    For Fourier reconstruction, we use rectangular filters for $H_\text{LPF}(d)$ and $H_\text{BPF}(d)$ as described in Eqs. (\ref{hlpf})--(\ref{hbpf}).
    We set the bandpass filter center to $d_0 = 36 \, \mu m$, corresponding to the carrier frequency, and the filter widths to $\Delta = 36 \, \mu m$.
    These filters are chosen to maintain equal spectral resolution in both channels.
    Note that the spectral resolution is $\Delta / 2$.

    To set the regularizer weight $\beta$, we first estimate the noise level of the detector by blocking light and calculating the power from Eq. (\ref{eqn:pn}).
    Following the description of SNR from Section \ref{SectionSNR}, we calculate the SNR to be greater than 40 dB for all samples.
    We choose $\beta=0.1$ since the SNR is high, using the guidance from Section \ref{SectionSNR}.

    To set the threshold $\tau$, we first note that the carrier frequency for our experimental data is $f_c = 36\, \mu m$.
    As described in Section \ref{sectionCS}, $\tau$ can be set close to the carrier frequency as one possible guideline.
    We set $\tau = 45 \, \mu m$, following this guideline.

    Figure \ref{fig:stokes} shows reconstructions for the four samples when the sample polarizer is oriented at $67.5^\circ$.
    For each sample, we compare ground truth with Fourier reconstruction and CCSP.
    We evaluate fit with ground truth using Eqs. (\ref{eqn:FRFit})--(\ref{eqn:CSFit}) and percent change from Fourier reconstruction to CCSP using Eq. (\ref{eqn:Delta}).
    Note the fits are displayed on the righthand side of each plot to quantify how well each algorithm performs.

    For sample 1 (LP) in Fig. \ref{fig:stokes1}, Fourier reconstruction displays artifacts at the edges of the spectrum, where the signal falls off, and noise results in oscillations in the Stokes parameters.
    In contrast, CCSP produces a more accurate, smoother reconstruction without the signal dropping at the edges.

    For sample 2 (Filter--LP) in Fig. \ref{fig:stokes2}, both CCSP and Fourier reconstruction capture the spectral variations in $S_0$, but Fourier reconstruction shows larger, non-physical noise in DOLP, $S_1 / S_0$, and $S_2 / S_0$.

    For sample 3 (Filter--LP--R) in Fig. \ref{fig:stokes3}, the quartz retarder creates a sinusoidal variation in DOLP.
    Fourier reconstruction fails to reconstruct DOLP correctly and displays non-physical oscillations in $S_2 / S_0$, but CCSP captures the sinusoidal DOLP, $S_0$, $S_1 / S_0$, and $S_2 / S_0$ more accurately.

    For sample 4 (LP--R) in Fig. \ref{fig:stokes4}, Fourier reconstruction doesn't accurately capture the sinusoidal DOLP variation and displays sinusoidal artifacts in $S_0$ and $S_2 / S_0$.
    In contrast, CCSP does not contain sinusoidal artifacts and results in more accurate values.

    Figure \ref{fig:fitangles} shows the fits over all orientations of the polarizer in the four samples.
    Note that Fig. \ref{fig:stokes} displays reconstructions at one orientation ($67.5^\circ$) of the polarizer as an example.
    We repeat these reconstructions over all angles and plot the fits in Fig. \ref{fig:fitangles}.
    We average the fits over all angles and display the averages on the righthand side of the plots.
    We also calculate the percent change of the average fit from Fourier reconstruction to CCSP, displayed as $\Delta_{\%}$ on the righthand side.
    For many parameters, CCSP offers a significant improvement; for example, $\Delta_{\%} = 61.4\%$ for $S_2 / S_0$ in sample 4 (LP--R).
    For some parameters, CCSP and Fourier reconstruction perform similarly; for example, FR fit = 77.1\% and CCSP fit = 79.4\% for $S_0$ in sample 3 (Filter--LP--R).
    Table \ref{StatisticsTable} summarizes the average fits for the four samples.

    These experiments demonstrate that CCSP produces more accurate reconstructions overall for a variety of samples.
    In particular, CCSP mitigates artifacts seen in Fourier reconstruction, as shown in Fig. \ref{fig:stokes}.
    These artifacts include signal falloff at spectral edges, noise sensitivity, nonphysical oscillations in birefringent samples, and bandwidth limitations caused by windowing.

\section{Conclusion}

We have presented a reconstruction method called compressed channeled spectropolarimetry (CCSP).
    In our proposed framework, reconstruction in channeled spectropolarimetry is an underdetermined problem, where we take $N$ measurements and solve for $3 N$ unknown Stokes parameters.
    We have formulated an optimization problem by creating a mathematical model of the channeled spectropolarimeter with inspiration from compressed sensing.
Our simulations and experiments have shown that CCSP produces more accurate reconstructions as tested over a variety of samples.
In particular, CCSP mitigates artifacts seen in Fourier reconstruction.
These artifacts include signal falloff at spectral edges, noise sensitivity, nonphysical oscillations in birefringent samples, and bandwidth limitations caused by windowing.
    By demonstrating more accurate reconstructions, we push performance to the native resolution of the sensor, allowing more information to be recovered from a single measurement of a channeled spectropolarimeter.

\section*{Acknowledgments}

Sandia National Laboratories is a multimission laboratory managed and operated by National Technology and Engineering Solutions of Sandia, LLC., a wholly owned subsidiary of Honeywell International, Inc., for the U.S. Department of Energy's National Nuclear Security Administration under contract DE-NA-0003525. The SAND number is SAND2017-7987 J.


\begin{thebibliography}{9}
    \bibitem{Kudenov11} M. W. Kudenov, M. J. Escuti, E. L. Dereniak, and K. Oka, ``White-light channeled imaging polarimeter using broadband polarization gratings,'' Appl. Opt. \textbf{50}(15),
2283--2293 (2011).
    \bibitem{Tyo06} J. S. Tyo, D. L. Goldstein, D. B. Chenault, and J. A. Shaw, ``Review of passive imaging polarimetry for remote sensing applications'', Appl. Opt. \textbf{45}, 5453--5469 (2006).
    \bibitem{Diner07} D. J. Diner, A. Davis, B. Hancock, G. Gutt, R. A. Chipman, and B. Cairns, ``Dual-photoelastic-modulator-based polarimetric imaging concept for aerosol remote sensing,'' Appl. Opt. \textbf{46}, 8428--8445 (2007).
    \bibitem{Boyer16} J. Boyer, J. C. Keresztes, W. Saeys, and J. Koshel, ``An automated imaging BRDF polarimeter for fruit quality inspection,'' Proc. SPIE \textbf{9948}, (2016).
    \bibitem{Peng12} B. Peng, T. Ding, and P. Wang, ``Propagation of polarized light through textile material,'' Appl. Opt. \textbf{51}, 6325-6334 (2012).
    \bibitem{Alenin14} A. S. Alenin and J. S. Tyo, ``Generalized channeled polarimetry,'' JOSA A \textbf{31}, 1013--1022 (2014).
    \bibitem{Lowenstern16} M. Lowenstern and M. W. Kudenov, ``Field deployable pushbroom hyperspectral imagining polarimeter,'' Proc. SPIE \textbf{9853}, (2016).
    \bibitem{Woodard16} E. R. Woodard and M. W. Kudenov, ``Spectrally resolved longitudinal spatial coherence inteferometry,'' Proc. SPIE \textbf{10198}, (2016).
    \bibitem{LaCasse15} C. F. LaCasse and O. G. Rodr\'{i}guez-Herrera and R. A. Chipman and J. S. Tyo, ``Spectral density response functions for modulated polarimeters,'' Appl. Opt. \textbf{54}, 9490--9499 (2015).
    \bibitem{Kudenov12} M. W. Kudenov and E. L. Dereniak, ``Compact real-time birefringent
imaging spectrometer,'' Opt. Express \textbf{20}(16), 17973--17986 (2012).
    \bibitem{Aspnes88} D. E. Aspnes, ``Analysis of Semiconductor Materials and Structures by
    Spectroellipsometry,'' Proc. SPIE \textbf{0946}, 84 (1988).
    \bibitem{Oka99} K. Oka and T. Kato, ``Spectroscopic polarimetry with a channeled spectrum,'' Opt.
Lett. \textbf{24}, 1475--1477 (1999).
    \bibitem{Chipman95} R. A. Chipman, ``Handbook of Optics,'' Polarimetry, 2nd ed. M. Bass ed. McGraw Hill, New York \textbf{2} (1995).
    \bibitem{Kudenov07} M. W. Kudenov, N. A. Hagen, E. L. Dereniak, and G. R. Gerhart, ``Fourier transform channeled spectropolarimetry in the MWIR,'' Opt. Express \textbf{15}, 12792--12805 (2007).
       \bibitem{Lee14} D. J. Lee and A. M. Weiner, ``Optical phase imaging using a synthetic aperture phase retrieval technique,'' Opt. Express \textbf{22}(8), 9380-9394 (2014).
       \bibitem{Lee15} D. J. Lee, K. Han, H. J. Lee, and A. M. Weiner, ``Synthetic aperture microscopy based on referenceless phase retrieval with an electrically tunable lens,'' Appl. Opt. \textbf{54}(17), 5346-5352 (2015).
       \bibitem{LaCasse11} C. F. LaCasse, R. A. Chipman, and J. S. Tyo, ``Band limited data
    reconstruction in modulated polarimeters,'' Opt. Express \textbf{19}(16), 14976--14989
    (2011).
\bibitem{Donoho06} D. L. Donoho, ``Compressed sensing,'' IEEE Transactions on Information Theory \textbf{52}, 1289--1306 (2006).
\bibitem{LeeSPIE16}  D. J. Lee, C. F. LaCasse, and J. M. Craven, ``Channeled spectropolarimetry using iterative reconstruction,'' Proc. SPIE \textbf{9853}, 98530V (2016).
\bibitem{LeeArxiv16} D. J. Lee, C. A. Bouman, and A. M. Weiner, ``Single Shot Digital Holography Using Iterative Reconstruction with Alternating Updates of Amplitude and Phase,'' http://www.arxiv.org/abs/1609.02978 (2016).
\bibitem{Boyd04} S. Boyd and L. Vandenberghe, \emph{Convex Optimization} (Cambridge University Press, 2004).
\bibitem{Wright97} S. Wright, \emph{Primal-Dual Interior-Point Methods} (Society for Industrial and Applied Mathematics, 1997).
\bibitem{Domahidi13} A. Domahidi, E. Chu, and S. Boyd, ``ECOS: An SOCP solver for embedded systems,'' in ``European Control Conference (ECC),'' (2013), pp. 3071-3076.
\end{thebibliography}
\end{document}